\documentclass[reqno]{amsart}
\usepackage{amsfonts}
\usepackage{amsmath}
\usepackage{hyperref} 

\newtheorem{theorem}{Theorem}[section]
\newtheorem{corollary}[theorem]{Corollary}
\newtheorem{proposition}[theorem]{Proposition}
\numberwithin{equation}{section}

\def\be{\begin{equation}}
\def\ee{\end{equation}}
\def\ba{\begin{eqnarray*}}
\def\ea{\end{eqnarray*}}
\def\bae{\begin{eqnarray}}
\def\eae{\end{eqnarray}}
\def\bc{\begin{center}}
\def\ec{\end{center}}

\newcommand{\Pf}{{\rm Pf}}
\newcommand{\tr}{{\rm tr}}
\newcommand{\e}{{\rm e}}
\newcommand{\diag}{{\rm diag}}

\begin{document}

\title[Bures measure and Cauchy two-matrix model]{Relating the Bures measure to the Cauchy two-matrix model}
\author{Peter J. Forrester} \address{Department of Mathematics and Statistics, The University of Melbourne, Victoria 3010, Australia}\email{p.forrester@ms.unimelb.edu.au}
\author{Mario Kieburg} \address{Fakult\"at f\"ur Physik, Universit\"at Bielefeld, Postfach 100131, D-33501 Bielefeld}\email{mkieburg@physik.uni-bielefeld.de}




\maketitle

\begin{abstract}
The Bures metric is a natural choice in measuring the distance of density operators representing states in quantum mechanics. In the past few years a random matrix ensemble and the corresponding joint probability density function of its eigenvalues was identified. Moreover a relation with the Cauchy two-matrix model was discovered but never thoroughly investigated, leaving open
in particular the following question: How are the kernels of the Pfaffian point process of the Bures random matrix ensemble related to the ones of the determinantal point process of the Cauchy two-matrix model and moreover, how can it be possible that a Pfaffian point process derives from a determinantal point process? We give a very explicit answer to this question. The aim of our work has a quite practical origin since the calculation of the level statistics of the Bures ensemble is highly mathematically involved while we know the statistics of the Cauchy two-matrix ensemble. Therefore we solve the whole level statistics of a density operator drawn from the Bures prior.
\end{abstract}

\section{Introduction}\label{intro}

The measuring and estimation of the distance between a density operator $\rho_{\rm exp}$ assumed after a finite number of experiments and the true quantum state $\rho_{\rm true}$ is an intrinsically hard task~\cite{schmied}. It becomes even harder since the metric on the set of quantum states is not uniquely determined by the standard conditions that a density operator $\rho$ has to be positive definite, Hermitian and the trace is equal to one ($\tr\, \rho=1$). Imposing additional conditions onto the metric restricts this ambiguity. The Bures metric~\cite{bures},
\begin{equation}\label{Bures-distance}
D(\rho_1,\rho_2)=\sqrt{2-2\tr\sqrt{\sqrt{\rho_1}\rho_2\sqrt{\rho_1}}}=\sqrt{2-2\tr\sqrt{\sqrt{\rho_2}\rho_1\sqrt{\rho_2}}},
\end{equation}
plays a distinguished role since it is the only metric which is also monotone, Fisher-adjusted, Fubini-Study-adjusted, and Riemannian, see Ref.~\cite{ZS03} for a clear explanation of these notions. One important application of the Bures metric is to define a geometric quantum discord measuring the strength of quantum correlations. Some recent works on this topic are Refs.~\cite{AFA13,BCFA14,ECO14,GTSQ14,HT14,OCCE14,PGST14,RGI14,SO14}.

The property that the Bures metric is Riemannian is particularly important for the application of random matrix theory. The Riemannian length element can be obtained by considering the distance between an $N\times N$ dimensional density operator $\rho$ with eigenvalues $z=\diag(z_1,\ldots,z_N)\in[0,1]^N$ and its infinitesimal neighbour $\rho+d\rho$ yielding~\cite{Hu92,Hu93}
\begin{equation}\label{length-element}
ds^2=[D(\rho,\rho+d\rho)]^2=\frac{1}{2}\sum_{i,j=1}^N\frac{d\rho_{ij}^2}{z_i+z_j}.
\end{equation}
The  joint probability density of the eigenvalues of $\rho$ is then~\cite{Hall}
\begin{equation}\label{fixed-trace}
p^{(N,a,{\rm fixed})}(z)\propto\delta\left(1-\sum_{j=1}^Nz_j\right)\prod_{j=1}^N z_j^{a} \prod_{1 \le i<j \le N}
{(z_j -z_i)^2 \over z_j + z_i},
\end{equation}
where the case $a=-1/2$  corresponds to  a full rank. When considering density operators of rank $M\leq N$ the exponent is $a=-1/2+N-M$, see Ref.~\cite{ZS03}. The latter case becomes important when measuring the separability of qubits and qutrits  (systems of two and three quantum states) on hyperareas in the set of quantum states~\cite{Slater04,Slater05}.

We remark that the joint probability density~\eqref{fixed-trace} can be interpreted as a log-gas with the pair-interaction $\exp[2{\rm ln}|z_i-z_j|-{\rm ln}|z_i+z_j|]$. This interpretation has been successfully applied in \cite{BorNad} to calculate the distribution of the purity $\Sigma_2=\tr \,\rho^2=\sum_{j=1}^N z_j^2$ in the limit of large $N$.

The authors of Ref.~\cite{OSZ10} showed that the ensemble of density operators of dimension $N\times N$ with rank $M\leq N$ distributed via the Bures measure can be generated by the random matrix\footnote{Note that in Ref.~\cite{OSZ10} the order of the product is the other way around since the non-zero eigenvalues do not depend on this order. However in their order the density operator has full rank because the generic zero modes are chopped off.}
\begin{eqnarray}\label{random-matrix}
 \rho=\frac{A(1+U)(1+U^\dagger)A^\dagger}{\tr(1+U)AA^\dagger(1+U^\dagger)},
\end{eqnarray}
where the complex $N\times M$ matrix $A$ is distributed by a Gaussian and the unitary $M\times M$ matrix $U$ by the Haar-measure on the unitary group ${\rm U}(M)$ with the prefactor $|{\det}(1+U)|^{2(N-M)}$. This approach connects the Bures measure with another topic in random matrix theory, namely product matrices. Quite recently there is a revival of interest in product matrices due to breakthroughs in the approach of asymptotic freeness and of integrable structures, with the latter applying at finite matrix dimension $N$ and finite number of matrices multiplied. This allowed the study 
of  the singular values and the eigenvalues 
 for finite matrix size as well as in particular limits for all three Dyson indices $\beta=1,2,4$ corresponding to real, complex and quaternion matrices, respectively, see Refs.~\cite{AKW13,AIK13,Fo13,Fo14,KS14,KZ13,NS14,ABKN14,KJ13,Zhang13,Strahov14,Burda13,BLS13,AB12,BNS12,BJLNS11,BJW09,Ipsen13,AIS14,PZ11,Neuschel13}. Even the convergence properties of Lyapunov exponents for a product of infinitely many matrices can now be studied in detail exhibiting interesting scaling limits \cite{Fo12,ABK14}. Additionally,  interesting relations to the distributions which have the combinatorial series of the Fu\ss-Catalan numbers, the Raney numbers and their generalizations as their moments \cite{ZS04,PZ11,FL14,ABK14,Neuschel13} have arisen. Exactly this relation was recently employed to calculate the level density for a generalized version of the Bures measure \cite{MNPZ14}.

One particular limit is on the local scale of the mean level spacing at the hard edge,  which is also known as the microscopic limit since it is the scale of the smallest eigenvalues around the origin. It was shown that product matrices show a different universal behaviour than the chiral Gaussian random matrix ensembles. The eigenvalues and singular values are no longer described by the Bessel kernel but by Meijer G-functions which seem to also hold for a larger class of random matrix ensembles \cite{AB12,ABKN14,Fo14,KS14,KZ13,NS14}. Exactly the same questions about the level statistics can be asked for density operators distributed via the Bures measure. Are those operators also in the universality class given by Meijer G-functions? Indeed this question is legitimate since another random matrix ensemble,  namely the Cauchy two-matrix model~\cite{BGS08},
 also exhibits a Meijer G-kernel at the hard edge~\cite{BGS12}, and is also known to involve products of
 random matrices  \cite{Fo14}. Interestingly, \cite{BGS08} contains  a relation between a particular kind of a Cauchy two-matrix model and the Bures ensemble. This relation was not worked out in detail since only the normalization constants of both ensembles were considered.   Thus the question is if the Cauchy two-matrix model and the Bures prior have more in common than the normalization. This was raised in \cite{BGS08}:
 \textit{``The relationship between the two model does not seem to go much further in the sense that there is
no direct and simple relationship between the correlation functions of the two models. It seems, however,
that some connection should be present and is worth exploring. We leave it as an open problem to
establish a connection between these two models on the level of the correlation functions."} Here we solve this open problem by expressing
 the correlation kernels for Bures measure in terms of those for the Cauchy two-matrix model. Thus we have an analytical formula for all eigenvalue statistics, and in particular
 for the eigenvalue density of the density operators  at finite matrix dimension $N$. 
 One consequence, to be explored in a subsequent work, is thus corresponding analytical formulae for moments of the eigenvalue density.
 Existing studies relating to this topic are restricted to numerical computations in the $N=3$ case~\cite{Slater12}, with a number of exact results conjectured.

The joint probability distribution of the eigenvalues we are considering is not exactly the one of the Bures measure~\eqref{fixed-trace} but the Fourier-Laplace transform of it,
\begin{equation}\label{1.1}
p^{(N,a,{\rm B})}(z)\propto\prod_{j=1}^N z_j^{a} \e^{-z_j} \prod_{1 \le i<j \le N}
{(z_j -z_i)^2 \over z_j + z_i},
\end{equation}
where $a > - 1$ and each $z_\nu \ge 0$. Despite this fact we call the random matrix ensemble corresponding to this distribution the Bures ensemble, too. It is exactly the distribution~\eqref{1.1} which is directly related to the Cauchy two-matrix ensemble. This relation is remarkable since the Cauchy two-matrix model is a determinantal point process corresponding to bi-orthogonal polynomials~\cite{BGS09} (all important facts will be recall in section~\ref{sec:Recall}) while the joint probability distribution~\eqref{1.1} forms a Pfaffian point process (see e.g.~\cite[Ch.~6]{Fo10}) due to Schur Pfaffian identity (see e.g.~\cite{IOTZ06}). As is well known from the study of the Hilbert-Schmidt measure in the context of quantum density matrices, the imposition of the Dirac delta function in mapping from the correlations for Eq.~\eqref{1.1} to those for the original Bures measure~\eqref{fixed-trace} requires coupling of the Pfaffian structure at finite $N$ to an auxiliary scaling variable multiplying each eigenvalue, which in turn is subject to a Fourier-Laplace transform (see e.g.~\cite[Eq.~(2.28)]{LZ10}). Although not pursued in the present work, again results for the Hilbert-Schmidt measure~\cite{LZ10} lead us to expect that the local spectral statistics should be the same due to universality. Furthermore, the joint probability density~\eqref{1.1} is interesting on its own since it can be also found in a completely different topic. In two-dimensional quantum gravity $p^{(N,a,{\rm B})}(z)$ is related to the ${\rm O}(N)$ vector model, see~\cite{Kos89}.

The determinantal point process of the Cauchy two-matrix model is recalled in section~\ref{sec:Recall}. Thereby we also calculate the average of ratios and products of characteristic polynomials of the two matrices which was not done before. This derivation is analogous to the computations in \cite{BDS,BorStr06,KG10,StrFyo}. On first sight it seems to be impossible to marry the kernels of the Bures ensemble with the Cauchy two-matrix model. We show in section~\ref{sec:Bures} that this is nonetheless possible because of the particular form of the joint probability density.  Pursing this approach we are able to express all kernels of the Bures ensemble in terms of the kernels for the Cauchy two-matrix model which were already derived in Refs.~\cite{BGS08,BGS09,BGS12}.

In section~\ref{sec:conclusio} we summarize and briefly discuss our results. In the appendices we present the details of our proofs.

\section{Recalling the Cauchy two-matrix model}\label{sec:Recall}

Since the Cauchy two-matrix model, introduced in Ref.~\cite{BGS08}, plays a crucial role we want to recall relevant known facts.
The general Cauchy two-matrix ensemble is a measure on two $N \times N$ positive definite Hermitian matrices $M_1, M_2$
distributed by
\begin{eqnarray}
P^{\rm(C)}(M_1,M_2)=\frac{\exp[-N(V_1(M_1) + V_2(M_2)+\tr\,{\rm ln}(M_1+M_2))]}{\int d[M_1] \int d[M_2]\exp[-N(V_1(M_1) + V_2(M_2))-N\tr\,{\rm ln}(M_1+M_2)]},\nonumber\\ \label{matrix-density-Cauchy-def}
\end{eqnarray}
where $d[M_{1}]$ and $d[M_{2}]$ are the products of differentials of the real independent matrix entries of $M_{1}$ and $M_{2}$, respectively. The functions $V_1$, $V_2$ are referred to as potentials. For the special choice $\exp[-NV_1(M)]= \det^a M \exp[- \tr \, M]$, $\exp[-NV_2(M)] = \det^b M$ $\exp[-\tr\, M]$,
the corresponding joint probability density of the eigenvalues of $M_1$ and $ M_2$ takes the form
\begin{equation}\label{jpdf-Cauchy-def}
p^{(N,a,b,{\rm C})}(x,y)\propto{\prod_{j=1}^N x_j^a \e^{-x_j} y_j^b \e^{-y_j}  \prod_{1 \le j < k \le N} (x_k - x_j)^2 (y_k - y_j)^2 \over
\prod_{j=1}^N \prod_{k=1}^N (x_j + y_k) }.
\end{equation}
 It enjoys a number of special integrability properties \cite{BGS12},
culminating in the explicit evaluation of the joint hard edge correlation function.

In subsection~\ref{subsec:Setting}, we introduce general partition functions  for the Cauchy two-matrix ensemble as the integral over products and ratios of characteristic polynomials weighted by the joint probability density~\eqref{jpdf-Cauchy-def}. Starting from these partition functions we recall the bi-orthogonal polynomials, their Cauchy transforms, and some other transforms employed in Ref.~\cite{BGS08,BGS09,BGS12}, in subsection~\ref{subsec:Polynomials}. These polynomials together with partition functions comprising of only  two characteristic polynomials (presented in subsection~\ref{subsec:Partition}) build the kernels of the determinantal point process which the Cauchy two-matrix model obeys. Eventually we  exhibit the explicit form of the $(k,l)$-point correlation function of the joint probability density~\eqref{jpdf-Cauchy-def} in subsection~\ref{subsec:Correlation}. When presenting these known results we briefly sketch their derivation to make our work self-contained, particularly introducing our notation and the normalization which slightly differs from the one in Ref.~\cite{BGS08,BGS09,BGS12}.

\subsection{Setting of the Cauchy two-matrix model}\label{subsec:Setting}

Generalised partition functions with ratios of characteristic polynomials play a crucial role when analyzing spectral statistics, see \cite{BDS,BorStr06,Kieburg12,KG10,KG10b,StrFyo} for the computations of those averages for some ensembles. This is also  true for the Cauchy two-matrix ensemble whose partition function may consist of four sets of characteristic polynomials
\begin{eqnarray}
Z^{(N,a,b,{\rm C})}_{k_1|l_1;k_2|l_2}(\kappa_1,\lambda_1;\kappa_2,\lambda_2)&:=&\frac{1}{(N!)^2}\int_{\mathbb R_+^{2N}}{ \Delta_N^2(x)  \Delta_N^2(y)\, d[x] d[y] \over \prod_{i,j=1}^{N} (x_i + y_j)}  \nonumber\\
  &&\hspace*{-1cm}\times\prod_{j=1}^{N} \left( x_j^ay_j^b\e^{-x_j-y_j}\frac{\prod_{i=1}^{l_1}(x_j-\lambda_{1,i})\prod_{i=1}^{l_2}(y_j-\lambda_{2,i})}{\prod_{i=1}^{k_1}(x_j-\kappa_{1,i})\prod_{i=1}^{k_2}(y_j-\kappa_{2,i})}\right).\nonumber\\
 \label{part-Cauchy-def}
\end{eqnarray}
Here the variables $\kappa_{1}=\diag(\kappa_{1,1},\ldots,\kappa_{1,k_1})$ and $\kappa_{2}=\diag(\kappa_{2,1},\ldots,\kappa_{2,k_2})$ do not lie on the positive real line while $\lambda_{1}=\diag(\lambda_{1,1},\ldots,\lambda_{1,l_1})$ and $\lambda_{2}=\diag(\lambda_{2,1},\ldots,\lambda_{2,l_2})$ can be arbitrary complex numbers.  Moreover we have used the Vandermonde determinant
 \begin{eqnarray}\label{Vand}
 \Delta_N(u) = \prod_{1 \le i < j \le N} (u_j - u_i)=\det[u_a^{b-1}]_{1\leq a,b,\leq N}
 \end{eqnarray}
which is one part of the Jacobian from diagonalizing the matrices $M_1$ and $M_2$. The notation of the indices of the partition function~\eqref{part-Cauchy-def} is reminiscent of the one used in the supersymmetry approach applied to random matrix theory, see Ref.~\cite{Berezin-book,KG10}, namely this kind of partition function is intimately related to supersymmetry. Moreover the prefactor $1/(N!)^2$ normalizes the partition function such that it is equivalent to one for an ordered set of variables $x_1\leq x_2\leq\ldots\leq x_N$ and $y_1\leq y_2\leq\ldots\leq y_N$. However such an ordering is often quite inconvenient for calculations so we consider the integral without an ordering. The measures $d[x]=dx_1 dx_2\cdots dx_N$ and $d[y]=dy_1 dy_2\cdots dy_N$ are the products of the differentials of the variables $x$ and $y$.

Let us consider some particular cases of these partition functions. The case $k_1=k_2=l_1=l_2=0$ yields the normalization constant of the joint probability density~\eqref{jpdf-Cauchy-def},
\begin{eqnarray}
Z^{(N,a,b,{\rm C})}_{0|0;0|0}&=&\det\left[\int_{\mathbb{R}_+^2}dxdy\frac{x^{a+i-1}y^{b+j-1}\e^{-x-y}}{x+y}\right]_{1\leq i,j\leq N}\nonumber\\
&=&\prod_{j=1}^{N} \frac{[(j-1)!]^2\Gamma[a+j]\Gamma[b+j]\Gamma[a+b+j]}{\Gamma[a+b+N+j]},\label{normalization-Cauchy}
\end{eqnarray}
see Ref.~\cite{BGS08}. This result can be obtained by applying Andr\'eief's integration theorem~\cite{Andr} and evaluating the determinant.

\subsection{Bi-orthogonal polynomials of the Cauchy two-matrix model}\label{subsec:Polynomials}{}

The two kinds of bi-orthogonal polynomials can be calculated with the help of these partition functions,
since they correspond to the averaged characteristic polynomials for the variables $\{x_i\}$ and $\{y_j\}$.
These are the cases $(k_1|l_1;k_2|l_2)=(0|1;0|0)$ and $(k_1|l_1;k_2|l_2)=(0|0;0|1)$ and yield the polynomials in monic normalization~\cite{Bor99}
\begin{eqnarray}
 p_n^{(a,b)}(x)&=&(-1)^n\frac{Z^{(n,a,b,{\rm C})}_{0|1;0|0}(x)}{Z^{(n,a,b,{\rm C})}_{0|0;0|0}}\nonumber\\
 &=&\frac{\det\left[\begin{array}{c|c} \displaystyle \int_{\mathbb{R}_+^2}dx'dy\frac{x'^{a+i-1}y^{b+j-1}\e^{-x'-y}}{x'+y} & x^{i-1} \end{array}\right]\underset{1\leq j\leq n}{\underset{1\leq i\leq n+1}{\ }}}{\displaystyle\det\left[\int_{\mathbb{R}_+^2}dxdy\frac{x^{a+i-1}y^{b+j-1}\e^{-x-y}}{x+y}\right]_{1\leq i,j\leq n}}\nonumber\\
 &&\hspace*{-0.7cm}=\sum_{j=0}^n (-1)^{n-j}\left(\begin{array}{c} n \\ j \end{array}\right)\frac{\Gamma(a+b+n+j+1)\Gamma(a+b+n+1)\Gamma(a+n+1)}{\Gamma(a+b+2n+1)\Gamma(a+b+j+1)\Gamma(a+j+1)}x^j\nonumber\\
 &=& (-1)^n\frac{(a+b+1)_n(a+1)_n}{(a+b+n+1)_n}\ _2F_2\left(\left.{ -n,\ a+b+n+1 \atop a+b+1,\ a+1 }\right|x\right),\nonumber\\ \label{polynomials-Cauchy-a}
 \end{eqnarray}
 and
 \begin{eqnarray}
 \widetilde{p}_n^{(a,b)}(y)&=&(-1)^n\frac{Z^{(n,a,b,{\rm C})}_{0|0;0|1}(y)}{Z^{(n,a,b,{\rm C})}_{0|0;0|0}}\nonumber\\
 &=&(-1)^n\frac{(a+b+1)_n(b+1)_n}{(a+b+n+1)_n}\ _2F_2\left(\left.{ -n,\ a+b+n+1 \atop a+b+1,\ b+1 }\right|y\right).\nonumber\\ \label{polynomials-Cauchy-b}
\end{eqnarray}
Both polynomials can be derived by applying a generalized version of Andr\'eief's integration theorem~\cite{KG10,Andr} and then expanding and evaluating the determinant. These steps are illustrated in lines two
and three of \eqref{polynomials-Cauchy-a}. This calculation was performed in Ref.~\cite{BGS12}.
The functions $ _pF_q$ are the generalized hypergeometric functions given by~\cite{GRJ-book}
\begin{eqnarray}\label{hypergeometric-def}
 _pF_q\left(\left.{ a_1,\cdots,a_p \atop b_1,\cdots,b_q }\right|z\right)=\sum_{j=0}^\infty\frac{\prod_{l=1}^p (a_l)_j}{\prod_{l=1}^q (b_l)_j}\frac{z^j}{j!}
\end{eqnarray}
with
\begin{equation}\label{Pochhammer}
 (a)_j=\prod_{l=0}^{j-1}(a+l)=\frac{\Gamma(a+j)}{\Gamma(a)}
\end{equation}
the Pochhammer symbol.
The polynomials~\eqref{polynomials-Cauchy-a} and \eqref{polynomials-Cauchy-b} are bi-orthogonal with respect to the weight $g^{(a,b,{\rm C})}(x,y)=x^ay^b\exp[-x-y]/(x+y)$,
\begin{eqnarray}
\int_{\mathbb{R}_+^2}dxdy \,g^{(a,b,{\rm C})}(x,y)p_n^{(a,b)}(x)\widetilde{p}_l^{(a,b)}(y)&=&\frac{Z_{0|0;0|0}^{(n+1,a,b,{\rm C})}}{Z_{0|0;0|0}^{(n,a,b,{\rm C})}}\delta_{nl}\nonumber\\
\hspace*{-0.3cm}&&\hspace*{-2.5cm}=\frac{[n!]^2\Gamma[a+n+1]\Gamma[b+n+1](\Gamma[a+b+n+1])^2}{\Gamma[a+b+2n+2]\Gamma[a+b+2n+1]}\delta_{nl}.\nonumber\\
\label{biorthogonality}
\end{eqnarray}
Another particularly helpful representation of the bi-orthogonal polynomials is in terms of Meijer G-functions~\cite{GRJ-book},
 \begin{eqnarray}
G_{p,q}^{m,n} \left( \left. {a_1,\dots, a_n;a_{n+1}, \ldots,a_p \atop b_1,\dots, b_m;b_{m+1}, \ldots,b_q} \right|z\right)=
\int_C\frac{ds}{2\pi\imath} {z^s\prod\limits_{j=1}^m \Gamma ( b_j - s) \prod\limits_{j=1}^n \Gamma(1 - a_j + s) \over
\prod\limits_{j=m+1}^q \Gamma (1-  b_j + s) \prod\limits_{j=n+1}^p \Gamma(a_j - s) }  ,\nonumber\\ \label{Meijer-G-def}
\end{eqnarray}
where the contour $C$ goes from $-\imath\infty$ to $+\imath\infty$ and lets the poles of $\Gamma(b_j-s)$ on the right side of the path while the poles of $\Gamma(1-a_j+s)$ are on the left side. Since each generalized hypergeometric function is a Meijer G-function the polynomials~\eqref{polynomials-Cauchy-a} and \eqref{polynomials-Cauchy-b} are
\begin{eqnarray}
 p_n^{(a,b)}(x)&=& (-1)^n \frac{n! \Gamma(a+b+n+1)\Gamma(a+n+1)}{\Gamma(a+b+2n+1)} G_{2,3}^{1,1} \left( \left. {-a-b-n;n+1 \atop 0;-a,-a-b} \right|x\right),\nonumber\\
 \widetilde{p}_n^{(a,b)}(y)&=&(-1)^n \frac{n! \Gamma(a+b+n+1)\Gamma(b+n+1)}{\Gamma(a+b+2n+1)} G_{2,3}^{1,1} \left( \left. {-a-b-n;n+1 \atop 0;-b,-a-b} \right|y\right),\nonumber\\ \label{polynomials-Cauchy-c}
\end{eqnarray}
cf. Ref.~\cite{BGS12}.

The Cauchy-transform of these polynomials frequently appear, too. They are proportional to the partition functions with the indices $(k_1|l_1;k_2|l_2)=(1|0;0|0)$ and $(k_1|l_1;k_2|l_2)=(0|0;1|0)$,
\begin{eqnarray}
 &&\mathcal{C}[\widetilde{p}]_n^{(a,b)}(x)=(-1)^n\frac{Z^{(n,a,b,{\rm C})}_{1|0;0|0}(x)}{Z^{(n,a,b,{\rm C})}_{0|0;0|0}}\nonumber\\
&=&\frac{\det\left[\begin{array}{c|c} \displaystyle\int_{\mathbb{R}_+^2}dx'dy'\frac{x'^{a+j-1}y'^{b+i-1}\e^{-x'-y'}}{x'+y'} & \displaystyle\int_{\mathbb{R}_+^2}dx'dy'\frac{x'^ay'^{b+i-1}\e^{-x'-y'}}{(x-x')(x'+y')} \end{array}\right]\underset{1\leq j\leq n-1}{\underset{1\leq i\leq n}{\ }}}{\displaystyle\det\left[\int_{\mathbb{R}_+^2}dx'dy'\frac{x'^{a+i-1}y'^{b+j-1}\e^{-x'-y'}}{x'+y'}\right]_{1\leq i,j\leq n}}\nonumber\\
&=&\frac{\Gamma[a+b+2n]}{[(n-1)!]^2\Gamma[a+n]\Gamma[b+n]\Gamma[a+b+n]}\int_{\mathbb{R}_+^2}dx'dy'\frac{x'^a y'^b\e^{-x'-y'}}{(x-x')(x'+y')}\widetilde{p}_{n-1}^{(a,b)}(y'),\nonumber\\ \label{Cpolynomials-Cauchy-a}
 \end{eqnarray}
 and
\begin{eqnarray} 
 &&\mathcal{C}[p]_n^{(a,b)}(y)=(-1)^n\frac{Z^{(n,a,b,{\rm C})}_{0|0;1|0}(y)}{Z^{(n,a,b,{\rm C})}_{0|0;0|0}}\nonumber\\
 &=&\frac{\Gamma[a+b+2n]}{[(n-1)!]^2\Gamma[a+n]\Gamma[b+n]\Gamma[a+b+n]}\int_{\mathbb{R}_+^2}dx'dy'\frac{x'^a y'^b\e^{-x'-y'}}{(y-y')(x'+y')}p_{n-1}^{(a,b)}(x'),\nonumber\\ \label{Cpolynomials-Cauchy-b}
 \end{eqnarray}
 with $x,y\notin\mathbb{R}_+$. In Ref.~\cite{BGS12} it was shown that the integral over $y'$ for $\mathcal{C}[\widetilde{p}]_n^{(a,b)}$ and $x'$ for $\mathcal{C}[p]_n^{(a,b)}$ is equal to a Meijer G-function, too, such that
\begin{eqnarray}
\mathcal{C}[\widetilde{p}]_n^{(a,b)}(x)&=&\frac{(-1)^{n}(a+b+2n-1)}{(n-1)!\Gamma[a+n]}\int_{\mathbb{R}_+}dx'\frac{x'^a}{x'-x}G_{2,3}^{2,1} \left( \left. {-a-n+1;n+b \atop 0,b;-a} \right|x'\right),\nonumber\\
 \mathcal{C}[p]_n^{(a,b)}(y)&=&\frac{(-1)^{n}(a+b+2n-1)}{(n-1)!\Gamma[b+n]}\int_{\mathbb{R}_+}dy'\frac{y'^b}{y'-y}G_{2,3}^{2,1} \left( \left. {-b-n+1;n+a \atop 0,a;-b}\right|y'\right).\nonumber\\ \label{Cpolynomials-Cauchy-c}
 \end{eqnarray}
 Also the remaining integrals can be performed by using the following four remarkable properties of Meijer G-functions~\cite{PBBM-book}
 {\small \begin{eqnarray*}
 z^\gamma G_{p,q}^{m,n} \left( \left. {a_1,\dots, a_n;a_{n+1}, \ldots,a_p \atop b_1,\dots, b_m;b_{m+1}, \ldots,b_q} \right|z\right)
  =G_{p,q}^{m,n} \left( \left. {a_1+\gamma,\dots, a_n+\gamma;a_{n+1}+\gamma, \ldots,a_p+\gamma \atop b_1+\gamma,\dots, b_m+\gamma;b_{m+1}+\gamma, \ldots,b_q+\gamma} \right|z\right),
  \end{eqnarray*}
  \begin{eqnarray*}
  G_{p,q}^{m,n} \left( \left. {a_1,\dots, a_n;a_{n+1}, \ldots,a_p \atop b_1,\dots, b_m;b_{m+1}, \ldots,b_q} \right|z\right)
  =G_{q,p}^{n,m} \left( \left. { 1-b_1,\dots, 1-b_m;1-b_{m+1}, \ldots,1-b_q \atop 1-a_1,\dots, 1-a_n;1-a_{n+1}, \ldots,1-a_p} \right|\frac{1}{z}\right),
  \end{eqnarray*}
  \begin{eqnarray*}
  \int_0^\infty \frac{dz'}{z'} G_{p,q}^{m,n} \left( \left. {a_1,\dots, a_n;a_{n+1}, \ldots,a_p \atop b_1,\dots, b_m;b_{m+1}, \ldots,b_q} \right|\frac{z}{z'}\right)G_{p',q'}^{m',n'} \left( \left. {a'_1,\dots, a'_{n'};a'_{n'+1}, \ldots,a'_{p'} \atop b'_1,\dots, b'_{m'};b'_{m'+1}, \ldots,b'_{q'}} \right|z'\right) \\
  =G_{p+p',q+q'}^{m+m',n+n'} \left( \left. {a_1,\dots,a_n,a'_1,\dots, a'_{n'};a_{n+1},\dots,a_p,a'_{n'+1}, \ldots,a'_{p'} \atop b_1,\dots,b_m,b'_1,\dots, b'_{m'};b_{m+1},\dots,b_q,b'_{m'+1}, \ldots,b'_{q'}} \right|z\right),\end{eqnarray*}
  \begin{equation}
  G_{p+1,q+1}^{m,n+1} \left( \left. {a_1,\dots, a_n,c;a_{n+1}, \ldots,a_p \atop b_1,\dots, b_m;b_{m+1}, \ldots,b_q,c} \right|z\right)
  =G_{p,q}^{m,n} \left( \left. {a_1,\dots, a_n;a_{n+1}, \ldots,a_p \atop b_1,\dots, b_m;b_{m+1}, \ldots,b_q} \right|z\right)\\ \label{Meijer-G-relations}
 \end{equation}}
 and the identification for a special case of a Meijer G-function
 \begin{equation}\label{Meijer-G-special-case-a}
 G_{1,1}^{1,1} \left( \left. {0 \atop 0} \right|z\right)=\frac{1}{1+z}.
 \end{equation}
 Then the Cauchy transforms of the polynomials are also Meijer G-functions
\begin{eqnarray}
\mathcal{C}[\widetilde{p}]_n^{(a,b)}(x)&=&\frac{(-1)^{n+1}(a+b+2n-1)}{(n-1)!\Gamma[a+n]}G_{2,3}^{3,1} \left( \left. {-n;n+a+b-1 \atop -1,a-1,a+b-1} \right|-x\right),\nonumber\\
 \mathcal{C}[p]_n^{(a,b)}(y)&=&\frac{(-1)^{n+1}(a+b+2n-1)}{(n-1)!\Gamma[b+n]}G_{2,3}^{3,1} \left( \left. {-n;n+a+b-1 \atop -1,b-1,a+b-1} \right|-y\right).\nonumber\\ \label{Cpolynomials-Cauchy-d}
 \end{eqnarray}
 
When replacing $x\to x+\imath\varepsilon$ and $y\to y+\imath\varepsilon$ with $x,y\in\mathbb{R}_+$ and taking the imaginary part of the Cauchy transform in the limit $\varepsilon\to0$ we recover the result of Ref.~\cite{BGS12},
\begin{eqnarray*}
&&\frac{1}{\pi}\lim_{\varepsilon\to 0}{\rm Im}\,\mathcal{C}[\widetilde{p}]_n^{(a,b)}(x+\imath\varepsilon)\nonumber\\
&&\hspace*{0.5cm}=-\frac{\Gamma[a+b+2n]}{[(n-1)!]^2\Gamma[a+n]\Gamma[b+n]\Gamma[a+b+n]}\int_{\mathbb{R}_+}dy'\frac{x^a y'^b\e^{-x-y'}}{x+y'}\widetilde{p}_{n-1}^{(a,b)}(y')\nonumber\\
&&\hspace*{0.5cm}=\frac{(-1)^{n}(a+b+2n-1)}{(n-1)!\Gamma[a+n]}x^aG_{2,3}^{2,1} \left( \left. {-a-n+1;n+b \atop 0,b;-a} \right|x\right),
\end{eqnarray*}
\begin{eqnarray}
 &&\frac{1}{\pi}\lim_{\varepsilon\to 0}{\rm Im}\,\mathcal{C}[p]_n^{(a,b)}(y+\imath\varepsilon)\nonumber\\
&&\hspace*{0.5cm}=-\frac{\Gamma[a+b+2n]}{[(n-1)!]^2\Gamma[a+n]\Gamma[b+n]\Gamma[a+b+n]}\int_{\mathbb{R}_+}dx'\frac{x'^a y^b\e^{-x'-y}}{x'+y}p_{n-1}^{(a,b)}(x')\nonumber\\
&&\hspace*{0.5cm}=\frac{(-1)^{n}(a+b+2n-1)}{(n-1)!\Gamma[b+n]}y^bG_{2,3}^{2,1} \left( \left. {-b-n+1;n+a \atop 0,a;-b} \right|y\right).\nonumber\\ \label{Cpolynomials-Cauchy-e}
 \end{eqnarray}
 We show in subsection~\ref{subsec:CauchytoBures} that the expressions in terms of Meijer G-functions will carry over to the Bures ensemble as well.

\subsection{Determinantal point process of the Cauchy two-matrix model}\label{subsec:Partition}

In general the partition function~\eqref{part-Cauchy-def} gives rise to a determinantal point process (see Refs.~\cite{Bor99,KG10} for general ensembles corresponding to bi-orthogonal polynomials) and can be expressed in terms of partition functions with one and two characteristic polynomials only. Assuming that $N+l_1-k_1=N+l_2-k_2=\widetilde{N}>1$, and $\kappa_{1},\kappa_2,\lambda_1,\lambda_2$ pairwise different then we have
\begin{eqnarray}
&&\hspace*{-0.5cm}Z^{(N,a,b,{\rm C})}_{k_1|l_1;k_2|l_2}(\kappa_1,\lambda_1;\kappa_2,\lambda_2)=\frac{(-1)^{k_1(k_1-1)/2+k_2(k_2-1)/2+l_1l_2}Z^{(\widetilde{N},a,b,{\rm C})}_{0|0;0|0}}{{\rm B}_{k_1|l_1}(\kappa_1;\lambda_1){\rm B}_{k_2|l_2}(\kappa_2;\lambda_2)}\label{Cauchy-det-identity}\\
&&\hspace*{-0.5cm}\times\det\left[\begin{array}{c|c}  \displaystyle\frac{Z^{(\widetilde{N}+1,a,b,{\rm C})}_{1|0;1|0}(\kappa_{1,i};\kappa_{2,j})}{Z^{(\widetilde{N},a,b,{\rm C})}_{0|0;0|0}} & \displaystyle\frac{1}{\kappa_{1,i}-\lambda_{1,j}}\frac{Z^{(\widetilde{N},a,b,{\rm C})}_{1|1;0|0}(\kappa_{1,i},\lambda_{1,j})}{Z^{(\widetilde{N},a,b,{\rm C})}_{0|0;0|0}} \\ \hline  \displaystyle \overset{\ }{\frac{1}{\kappa_{2,j}-\lambda_{2,i}}\frac{Z^{(\widetilde{N},a,b,{\rm C})}_{0|0;1|1}(\kappa_{2,j},\lambda_{2,i})}{Z^{(\widetilde{N},a,b,{\rm C})}_{0|0;0|0}}} & \displaystyle -\frac{Z^{(\widetilde{N}-1,a,b,{\rm C})}_{0|1;0|1}(\lambda_{1,j};\lambda_{2,i})}{Z^{(\widetilde{N},a,b,{\rm C})}_{0|0;0|0}}  \end{array}\right],\nonumber
\end{eqnarray}
where the indices are $1\leq i\leq k_1$ in the first rows and  $1\leq i\leq l_2$ in the last rows and $1\leq j\leq k_2$ in the first columns and $1\leq j\leq l_1$ in the last ones. Recall that $\kappa_{1,j},\kappa_{2,j}\in\mathbb{C}\setminus\mathbb{R}_+$ for each $j=1,2,\ldots$ We employed the mixed Cauchy-Vandermonde determinant~\cite{BF94,KG10}
\begin{eqnarray}\label{Ber-main}
{\rm B}_{k|l}(\kappa;\lambda)&=&\frac{\Delta_k(\kappa)\Delta_l(\lambda)}{\prod_{i=1}^k\prod_{j=1}^l(\kappa_i-\lambda_j)}\nonumber\\
&&\hspace*{-1cm}=\left\{\begin{array}{cl}(-1)^{l(l-1)/2}\det\left[\begin{array}{cc} \displaystyle\left\{\frac{1}{\kappa_a-\lambda_b}\right\}\underset{1\leq b\leq l}{\underset{1\leq a\leq k}{}} & \displaystyle\left\{\kappa_a^{b-1}\right\}\underset{1\leq b\leq k-l}{\underset{1\leq a\leq k}{}} \end{array}\right], & k\geq l, \\ (-1)^{k(k-1)/2}\det\left[\begin{array}{cc} \displaystyle\left\{\lambda_a^{b-1}\right\}\underset{1\leq b\leq l-k}{\underset{1\leq a\leq l}{}} & \displaystyle\left\{\frac{1}{\kappa_b-\lambda_a}\right\}\underset{1\leq b\leq k}{\underset{1\leq a\leq l}{}} \end{array}\right], & k\leq l. \end{array}\right.\nonumber\\
\end{eqnarray}
This determinant plays a crucial role in the theory of supermatrices~\cite{Berezin-book}.

We underline that the case $k_1=l_1=k$ and $k_2=l_2=l$ is important when calculating the $(k,l)$-point correlation function of the Cauchy two-matrix model, see subsection~\ref{subsec:Correlation}. Another  important case of the partition function  is  when $k_1=k_2=k$ and $l_1=l_2=l$ which is needed in subsection~\ref{subsec:CauchytoBures}  to invert the relation between the Bures ensemble and the Cauchy two-matrix ensemble.

The general case of $k_1$, $k_2$, $l_1$ and $l_2$ arbitrary can be obtained by sending some of the variables $\kappa_{1},\kappa_2,\lambda_1,\lambda_2$ to infinity. Then one  also finds the bi-orthogonal polynomials~\eqref{polynomials-Cauchy-a} and \eqref{polynomials-Cauchy-b} and their Cauchy transform~\eqref{Cpolynomials-Cauchy-a} and \eqref{Cpolynomials-Cauchy-b} in  the determinant after applying l'Hospital's  rule and making use of the skew-symmetry of the determinant under permutation of rows and columns.

Let us look at the two-point partition functions in the kernels of the determinant~\eqref{Cauchy-det-identity} in detail. All four kernels can be expressed in terms of integrals over the bi-orthogonal polynomials and their Cauchy-transform and thus in terms of Meijer G-functions,
\begin{eqnarray}
 &&\frac{Z^{(N-1,a,b,{\rm C})}_{0|1;0|1}(\lambda_{1};\lambda_{2})}{Z^{(N,a,b,{\rm C})}_{0|0;0|0}}=-{\small\frac{\det\left[\begin{array}{c|c} \displaystyle \int_{\mathbb{R}_+^2}dxdy\frac{x^{a+i-1}y^{b+j-1}\e^{-x-y}}{x+y} & \lambda_1^{i-1} \\ \hline \lambda_2^{j-1} & 0 \end{array}\right]_{1\leq i,j\leq N}}{\det\left[\int_{\mathbb{R}_+^2}dxdy\frac{x^{a+i-1}y^{b+j-1}\e^{-x-y}}{x+y}\right]_{1\leq i,j\leq N}}}\nonumber\\ 
 &=&\sum_{i,j=0}^{N-1}\frac{\Gamma[a+b+i+N+1](-\lambda_1)^{i}}{i!(N-i-1)!\Gamma[a+b+i+1]\Gamma[a+i+1]}\nonumber\\
 &&\times\frac{\Gamma[a+b+j+N+1](-\lambda_2)^{j}}{j!(N-j-1)!\Gamma[a+b+j+1]\Gamma[b+j+1]}\frac{1}{a+b+j+i+1}\nonumber\\
 &=&\int_0^1dt t^{a+b}G_{2,3}^{1,1} \left( \left. {-N-a-b;N \atop 0;-a,-a-b} \right|t\lambda_1\right)G_{2,3}^{1,1} \left( \left. {-N-a-b;N \atop 0;-b,-a-b} \right|t\lambda_2\right)\nonumber\\
 \label{two-point-Cauchy-a}
\end{eqnarray}
for the average of two characteristic polynomials in the numerator (cf. Ref.~\cite{BGS12}),
\begin{eqnarray}
 &&\frac{1}{\kappa-\lambda}\frac{Z^{(N,a,b,{\rm C})}_{1|1;0|0}(\kappa,\lambda)}{Z^{(N,a,b,{\rm C})}_{0|0;0|0}}= {\small\frac{\det\left[\begin{array}{c|c} \displaystyle \int_{\mathbb{R}_+^2}dxdy\frac{x^{a+i-1}y^{b+j-1}\e^{-x-y}}{x+y} & \lambda^{i-1} \\ \hline \displaystyle\int_{\mathbb{R}_+^2}dxdy\frac{x^{a}y^{b+j-1}\e^{-x-y}}{(\kappa-x)(x+y)} & \displaystyle\frac{1}{\kappa-\lambda} \end{array}\right]_{1\leq i,j\leq N}}{\det\left[\int_{\mathbb{R}_+^2}dxdy\frac{x^{a+i-1}y^{b+j-1}\e^{-x-y}}{x+y}\right]_{1\leq i,j\leq N}}}\nonumber\\ 
 &=&\frac{1}{\kappa-\lambda}-\int_{\mathbb{R}_+^2}dxdy\frac{x^{a}y^{b}\e^{-x-y}}{(\kappa-x)(x+y)}\frac{Z^{(N-1,a,b,{\rm C})}_{0|1;0|1}(\lambda;y)}{Z^{(N,a,b,{\rm C})}_{0|0;0|0}}\nonumber\\
 &=&\frac{1}{\kappa-\lambda}+\int_0^1dt G_{2,3}^{1,1} \left( \left. {-N-a-b;N \atop 0;-a,-a-b} \right|t\lambda\right)G_{2,3}^{3,1} \left( \left. {-N;N+a+b \atop 0,a,a+b} \right|-t\kappa\right)\nonumber\\
 \label{two-point-Cauchy-b}
\end{eqnarray}
and analogously
\begin{eqnarray}
 &&\frac{1}{\kappa-\lambda}\frac{Z^{(N,a,b,{\rm C})}_{0|0;1|1}(\kappa,\lambda)}{Z^{(N,a,b,{\rm C})}_{0|0;0|0}}\nonumber\\
 &=&\frac{1}{\kappa-\lambda}+\int_0^1dt G_{2,3}^{3,1} \left( \left. {-N;N+a+b \atop 0,b,a+b} \right|-t\kappa\right)G_{2,3}^{1,1} \left( \left. {-N-a-b;N \atop 0;-b,-a-b} \right|t\lambda\right)\nonumber\\
 \label{two-point-Cauchy-c}
\end{eqnarray}
for the partition function with one characteristic polynomial in the numerator and one in the denominator, and
\begin{eqnarray}
 &&\frac{Z^{(N+1,a,b,{\rm C})}_{1|0;1|0}(\kappa_{1};\kappa_{2})}{Z^{(N,a,b,{\rm C})}_{0|0;0|0}}\nonumber\\
 &&\hspace*{-0.6cm}={\small\frac{\det\left[\begin{array}{c|c} \displaystyle \int_{\mathbb{R}_+^2}dxdy\frac{x^{a+i-1}y^{b+j-1}\e^{-x-y}}{x+y} & \displaystyle\int_{\mathbb{R}_+^2}dxdy\frac{x^{a+i-1}y^{b}\e^{-x-y}}{(\kappa_2-y)(x+y)}  \\ \hline \displaystyle\int_{\mathbb{R}_+^2}dxdy\frac{x^{a}y^{b+j-1}\e^{-x-y}}{(\kappa_1-x)(x+y)}  & \displaystyle\int_{\mathbb{R}_+^2}dxdy\frac{x^{a}y^{b}\e^{-x-y}}{(\kappa_1-x)(\kappa_2-y)(x+y)}  \end{array}\right]_{1\leq i,j\leq N}}{\det\left[\int_{\mathbb{R}_+^2}dxdy\frac{x^{a+i-1}y^{b+j-1}\e^{-x-y}}{x+y}\right]_{1\leq i,j\leq N}}}\nonumber\\ 
 &&\hspace*{-0.6cm}=\int_{\mathbb{R}_+^2}dxdy\frac{x^{a}y^{b}\e^{-x-y}}{(\kappa_1-x)(\kappa_2-y)(x+y)}-\sum_{j=0}^{N-1}(a+b+2j+1)\nonumber\\
 &&\times G_{2,3}^{3,1} \left( \left. {-j-1;j+a+b \atop -1,a-1,a+b-1} \right|-\kappa_1\right)G_{2,3}^{3,1} \left( \left. {-j-1;j+a+b-1 \atop -1,b-1,a+b-1} \right|-\kappa_2\right)\nonumber\\
 &&\hspace*{-0.6cm}=\int_{\mathbb{R}_+^2}dxdy\frac{x^{a}y^{b}\e^{-x-y}}{(\kappa_1-x)(\kappa_2-y)(x+y)}-(-1)^{a+b}\kappa_1^a\kappa_2^b\nonumber\\
 &&\times\int_0^1dt\left[ G_{2,3}^{3,1} \left( \left. {-a-N;N+b \atop 0,-a,b} \right|-t\kappa_1\right)G_{2,3}^{3,1} \left( \left. {-b-N;N+a \atop 0,-b,a} \right|-t\kappa_2\right)\right.\nonumber\\
 &&\left.-G_{2,3}^{3,1} \left( \left. {-a;b \atop 0,-a,b} \right|-t\kappa_1\right)G_{2,3}^{3,1} \left( \left. {-b;a \atop 0,-b,a} \right|-t\kappa_2\right)\right]\nonumber\\
 \label{two-point-Cauchy-d}
\end{eqnarray}
for two characteristic polynomials in the denominator. The last term can be derived by writing the Meijer G-function as contour integrals, see Eq.~\eqref{Meijer-G-def}, and employing the identity
\begin{eqnarray}
&&\sum_{j=0}^{N-1}(a+b+2j+1)\frac{\Gamma[j+u+2]\Gamma[j+v+2]}{\Gamma[j+a+b-u]\Gamma[j+a+b-v]}=\frac{1}{3-a-b+u+v}\nonumber\\
&&\hspace*{-0.7cm}\times\left[\frac{\Gamma[N+u+2]\Gamma[N+v+2]}{\Gamma[N+a+b-u-1]\Gamma[N+a+b-v-1]}-\frac{\Gamma[u+2]\Gamma[v+2]}{\Gamma[a+b-u-1]\Gamma[a+b-v-1]}\right],\nonumber\\
\label{sum-identity}
\end{eqnarray}
which is based on Lemma 4.1 of Ref.~\cite{BGS12} and can be proven by taking the difference of the sum for $N=k$ and $N=k-1$ and showing that both sides are the same. For the other results we used the same integral identities~\eqref{Meijer-G-relations} as for the polynomials.

We emphasize that the results~(\ref{two-point-Cauchy-a}-\ref{two-point-Cauchy-d}) are similar to but not exactly the same as the ones in Ref.~\cite{BGS12} where the Cauchy-transforms were not calculated. The transforms presented in Ref.~\cite{BGS12}  can be found by choosing $\kappa_j\to\kappa_j+L_j\imath \varepsilon$ ($j=1,2$) in the limit $\varepsilon\to 0$ with $\kappa_j\in\mathbb{R}_+$ and $L_j=\pm1$. Taking the differences of the results for $L_1=+1$ and $L_1=-1$ and the same for for $L_2$ we find
\begin{eqnarray}
 \lefteqn{-\frac{1}{2\pi\imath}\sum_{L_1=\pm1}\frac{L_1}{\kappa_1+L_1\imath\varepsilon-\lambda}\frac{Z^{(N,a,b,{\rm C})}_{1|1;0|0}(\kappa_1+L_1\imath\varepsilon,\lambda)}{Z^{(N,a,b,{\rm C})}_{0|0;0|0}} }\nonumber\\
 &&=\delta(\kappa_1-\lambda)-\kappa_1^{a+b}\int_0^1dt t^{a+b}G_{2,3}^{1,1} \left( \left. {-N-a-b;N \atop 0;-a,-a-b} \right|t\lambda\right)\nonumber\\
 &&\quad \times G_{2,2}^{3,1} \left( \left. {-N-a-b;N \atop 0,-b;-a-b} \right|t\kappa_1\right),\label{two-point-Cauchy-e}
\end{eqnarray}
and
\begin{eqnarray}
 \lefteqn{-\frac{1}{2\pi\imath}\sum_{L_2=\pm1}\frac{L_1}{\kappa_2+L_2\imath\varepsilon-\lambda}\frac{Z^{(N,a,b,{\rm C})}_{0|0;1|1}(\kappa_2+L_2\imath\varepsilon,\lambda)}{Z^{(N,a,b,{\rm C})}_{0|0;0|0}}}\nonumber\\
 &&= \delta(\kappa_2-\lambda)-\kappa_2^{a+b}\int_0^1dt t^{a+b}G_{2,2}^{3,1} \left( \left. {-N-a-b;N \atop 0,-a;-a-b} \right|t\kappa_2\right)\nonumber\\
 &&\quad \times G_{2,3}^{1,1} \left( \left. {-N-a-b;N \atop 0;-b,-a-b} \right|t\lambda\right),\label{two-point-Cauchy-f}
\end{eqnarray}
for the two kernels of the off-diagonal blocks and
\begin{eqnarray}
 &&\lefteqn{-\frac{1}{(2\pi)^2}\sum_{L_1,L_2=\pm1}L_1L_2\frac{Z^{(N+1,a,b,{\rm C})}_{1|0;1|0}(\kappa_{1}+L_1\imath\varepsilon;\kappa_{2}+L_2\imath\varepsilon)}{Z^{(N,a,b,{\rm C})}_{0|0;0|0}}}\nonumber\\
 &&=\frac{\kappa_1^{a}\kappa_2^{b}\e^{-\kappa_1-\kappa_2}}{\kappa_1+\kappa_2}-\kappa_1^{a}\kappa_2^{b}\sum_{j=0}^{N-1}(a+b+2j+1)\nonumber\\
 &&\quad \times G_{2,3}^{2,1} \left( \left. {-a-j;j+b+1 \atop 0,b;-a} \right|\kappa_1\right)G_{2,3}^{2,1} \left( \left. {-b-j;j+a+1 \atop 0,a;-b} \right|\kappa_2\right)\nonumber\\
 &&=\frac{\kappa_1^{a}\kappa_2^{b}}{\kappa_1+\kappa_2}-\kappa_1^{a}\kappa_2^{b}\nonumber\\
 &&\quad \times \int_0^1dt G_{2,3}^{2,1} \left( \left. {-a-N;N+b \atop 0,b;-a} \right|t\kappa_1\right)G_{2,3}^{2,1} \left( \left. {-b-N;N+a \atop 0,a;-b} \right|t\kappa_2\right)\nonumber\\
 \label{two-point-Cauchy-g}
\end{eqnarray}
which is up to the prefactor $-\kappa_1^a\kappa_2^b$ the kernel $K_{11}^{(N)}$ in Ref.~\cite{BGS12}. To obtain the equality of Eq.~\eqref{two-point-Cauchy-g} we have to notice that the second term of Eq.~\eqref{sum-identity} yields the Meijer G-function $G_{0,1}^{1,0}\left(\left.{- \atop 0;-}\right|\kappa_1 t\right)=e^{-\kappa_1t}$ and the same for $\kappa_2$. The integral over $t$ yields two terms such that it is correct that the first term of the second equality in Eq.~\eqref{two-point-Cauchy-g} does not contain an exponential function.

\subsection{Correlation functions of the Cauchy two-matrix model}\label{subsec:Correlation}

We now come to the eigenvalue correlation functions of the Cauchy two-matrix model after 
having recalled the results for the partition functions. Here, we emphasize that there are two definitions of the $(k,l)$-point correlation function at the positions $x=(x_1,\ldots,x_k)\in\mathbb{R}_+^k$ and $y=(y_1,\ldots,y_l)\in\mathbb{R}_+^l$, namely (see Refs.~\cite{Mehta-book,Fo10,Gernot-book} and references therein)
\begin{eqnarray}
\widehat{R}_{k,l}^{(N,a,b,{\rm C})}(x;y)&:=&\frac{1}{Z_{0|0;0|0}^{(N,a,b,{\rm C})}}\frac{1}{(N!)^2}\int_{\mathbb R_+^{2N}}{ \Delta_N^2(x')  \Delta_N^2(y')\, d[x'] d[y'] \over \prod_{i,j=1}^{N} (x'_i + y'_j)} \nonumber\\
  &&\hspace*{-2cm}\times\prod_{j=1}^{N} \left( {x'_j}^a{y'_j}^b\e^{-x'_j-y'_j}\right) \prod_{j=1}^k\left(\frac{1}{N}\sum_{i=1}^N\delta(x_j-x'_i)\right)\prod_{j=1}^l\left(\frac{1}{N}\sum_{i=1}^N\delta(y_j-y'_i)\right)\nonumber\\
  &=&\frac{1}{Z_{0|0;0|0}^{(N,a,b,{\rm C})}}\lim_{\varepsilon\to0}\sum_{L_j,L'_i=\pm}\prod_{j=1}^k\left(\frac{L_j}{2\pi\imath N}\frac{\partial}{\partial \widetilde{x}_j}\right)\prod_{i=1}^l\left(\frac{L_i}{2\pi\imath N}\frac{\partial}{\partial \widetilde{y}_i}\right)\nonumber\\
  &&\hspace*{-2cm}\times Z_{k|k;l|l}^{(N,a,b,{\rm C})}(\widetilde{x}+\imath L\varepsilon,x;\widetilde{y}+\imath L'\varepsilon,y)\biggl|_{\widetilde{x}=x,\widetilde{y}=y},\nonumber\\
 \label{correl-Cauchy-def}
\end{eqnarray}
where $L=(L_1,\ldots,L_k)$ and $L'=(L'_1,\ldots,L'_l)$ and
\begin{eqnarray}
R_{k,l}^{(N,a,b,{\rm C})}(x;y)&:=&\frac{1}{Z_{0|0;0|0}^{(N,a,b,{\rm C})}}\frac{1}{(N!)^2}\int_{\mathbb R_+^{2N-k-l}}\prod_{j=k+1}^Ndx_j\prod_{j=l+1}^Ndy_j{ \Delta_N^2(x)  \Delta_N^2(y) \over \prod_{i,j=1}^{N} (x_i + y_j)}\nonumber\\
 &&\times \prod_{j=1}^{N} \left( x_j^ay_j^b\e^{-x_j-y_j}\right). \label{conn-correl-Cauchy-def}
\end{eqnarray}
These two definitions are not equivalent. Nonetheless they are related in a simple way. The correlation function $\widehat{R}_{k,l}^{(N,a,b,{\rm C})}(x;y)$ consists not only of the correlation function $R_{k,l}^{(N,a,b,{\rm C})}(x;y)$ but also of the lower order terms like $R_{k-1,l}^{(N,a,b,{\rm C})}(x_1,\ldots,$ $x_{k-1};y)$ or $R_{k,l-1}^{(N,a,b,{\rm C})}(x;y_1,\ldots,y_{l-1})$. The reason is that $\widehat{R}_{k,l}^{(N,a,b,{\rm C})}(x;y)$ comprises ``self-energy'' terms proportional to Dirac delta-functions as $\delta(x_i-x_j)$ with $i\neq j$. Omitting these  ``self-energy''  terms we find $R_{k,l}^{(N,a,b,{\rm C})}(x;y)$, i.e.
\begin{equation}\label{connection-Cauchy-def}
\widehat{R}_{k,l}^{(N,a,b,{\rm C})}(x;y)=\frac{(N!)^2}{(N-k)!(N-l)!N^{k+l}}R_{k,l}^{(N,a,b,{\rm C})}(x;y)+{\rm lower\ order\ terms}.
\end{equation}

We remark that the definition for $\widehat{R}_{k,l}^{(N,a,b,{\rm C})}$ is based on the partition function~\eqref{Cauchy-det-identity} with $k_1=l_1=k$ and $k_2=l_2=l$ for which we already calculated a simplified determinantal expression in terms of two-point partition functions. In this expression we can easily perform the differentiations in $x$ and $y$ which essentially acts on the prefactor in front of the determinant only, since it vanishes at $\widetilde{x}+\imath L\varepsilon=x$ and $\widetilde{y}+\imath L'\varepsilon=y$. Only for the diagonal elements do we have to differentiate the kernel which yields one point functions. Therefore the result is
\begin{eqnarray}
&&\widehat{R}_{k,l}^{(N,a,b,{\rm C})}(x;y)\nonumber\\
&=&\det\left[\begin{array}{c|c} K_{01}^{(N,a,b,{\rm C})}(x_i,x_j) & K_{11}^{(N,a,b,{\rm C})}(x_i;y_j) \\ \hline K_{00}^{(N,a,b,{\rm C})}(x_j;y_i) & K_{10}^{(N,a,b,{\rm C})}(y_i,y_j) \end{array}\right]+{\rm lower\ order\ terms},\nonumber\\
 \label{correl-Cauchy-result}
\end{eqnarray}
where the indices take the same values as in Eq.~\eqref{Cauchy-det-identity}. The kernels are
\begin{eqnarray}
K_{11}^{(N,a,b,{\rm C})}(x_i;y_j)&=&-\frac{x_i^{a}y_j^{b}}{x_i+y_j}+x_i^{a}y_j^{b}\int_0^1dt G_{2,3}^{2,1} \left( \left. {-a-N;N+b \atop 0,b;-a} \right|tx_i\right)\nonumber\\
 &&\times G_{2,3}^{2,1} \left( \left. {-b-N;N+a \atop 0,a;-b} \right|ty_j\right),\nonumber\\
K_{01}^{(N,a,b,{\rm C})}(x_i,x_j)&=&x_i^{a+b}\int_0^1dt t^{a+b}G_{2,3}^{1,1} \left( \left. {-N-a-b;N \atop 0;-a,-a-b} \right|tx_j\right)\nonumber\\
 &&\times G_{2,3}^{2,1} \left( \left. {-N-a-b;N \atop 0,-b;-a-b} \right|tx_i\right),\nonumber\\
K_{10}^{(N,a,b,{\rm C})}(y_i,y_j)&=&y_j^{a+b}\int_0^1dt t^{a+b}G_{2,3}^{2,1} \left( \left. {-N-a-b;N \atop 0,-a;-a-b} \right|ty_j\right)\nonumber\\
 &&\times G_{2,3}^{1,1} \left( \left. {-N-a-b;N \atop 0;-b,-a-b} \right|ty_i\right),\nonumber\\
K_{00}^{(N,a,b,{\rm C})}(x_j;y_i) &=&\int_0^1dt t^{a+b}G_{2,3}^{1,1} \left( \left. {-N-a-b;N \atop 0;-a,-a-b} \right|tx_j\right)\nonumber\\
&&\times G_{2,3}^{1,1} \left( \left. {-N-a-b;N \atop 0;-b,-a-b} \right|ty_i\right).\nonumber\\
\label{kernel-Cauchy}
\end{eqnarray}
From this result one can read off the $(k,l)$-point correlation function without ``self-energy'' terms,
\begin{eqnarray}
R_{k,l}^{(N,a,b,{\rm C})}(x;y)=\frac{(N-k)!(N-l)!N^{k+l}}{(N!)^2}\det\left[\begin{array}{c|c} K_{01}^{(N,a,b,{\rm C})}(x_i,x_j) & K_{11}^{(N,a,b,{\rm C})}(x_i;y_j) \\ \hline K_{00}^{(N,a,b,{\rm C})}(x_j;y_i) & K_{10}^{(N,a,b,{\rm C})}(y_i,y_j) \end{array}\right],\nonumber\\
 \label{con-correl-Cauchy-result}
\end{eqnarray}
which is the result of Ref.~\cite{BGS12}. Note that our $k$-point functions are normalized, i.e. $\int d[x]d[y]R_{k,l}^{(N,a,b,{\rm C})}(x;y)=1$. In the particular case of the two kinds of level densities this result reads
\begin{eqnarray}
R_{1,0}^{(N,a,b,{\rm C})}(x)&=&K_{01}^{(N,a,b,{\rm C})}(x,x)\label{level-density-Cauchy-a}\\
&=&\int_0^1dt G_{2,3}^{1,1} \left( \left. {-N;N+a+b \atop a+b;0,b} \right|tx\right)G_{2,3}^{2,1} \left( \left. {-N-a-b;N \atop 0,-b;-a-b} \right|tx\right)\nonumber
\end{eqnarray}
and
\begin{eqnarray}
R_{0,1}^{(N,a,b,{\rm C})}(y)&=&K_{10}^{(N,a,b,{\rm C})}(y,y)\label{level-density-Cauchy-b}\\
&=&\int_0^1dt G_{2,3}^{1,1} \left( \left. {-N;N+a+b \atop a+b;0,a} \right|ty\right)G_{2,3}^{2,1} \left( \left. {-N-a-b;N \atop 0,-a;-a-b} \right|ty\right),\nonumber
\end{eqnarray}
which shall conclude this section. Due to the symmetry of the joint probability density~\eqref{jpdf-Cauchy-def} the spectral statistics is invariant by exchanging $\{x_i\}\leftrightarrow \{y_i\}$ and $a\leftrightarrow b$ which in particular is reflected in the level densities $R_{1,0}^{(N,a,b,{\rm C})}$ and $R_{0,1}^{(N,a,b,{\rm C})}$. Now we are well-prepared for calculating the eigenvalue statistics of the Bures ensemble.

\section{Relationship between Bures and Cauchy two-matrix ensemble}\label{sec:Bures}

We aim at two things in this section. First we want to work along the same ideas and calculations as we have done it for the Cauchy two-matrix ensemble in section~\ref{sec:Recall}. Hence we start with the partition function
\begin{equation}
Z^{(N,a,{\rm B})}_{k|l}(\kappa,\lambda)=\frac{1}{N!}\int_{\mathbb R_+^{N}}d[z]\frac{ \Delta_N^2(z) }{\prod_{1 \le i < j \le N}(z_i + z_j)}\prod\limits_{j=1}^N z_j^a\e^{-z_j}\frac{\prod_{i=1}^l(z_j-\lambda_i)}{\prod_{i=1}^k(z_j-\kappa_i)}\label{part-Bures-def}
\end{equation}
and express every other quantity in terms of this including the normalization constant, the skew-orthogonal polynomials, and the $k$-point correlation function. Thereby for the latter we again choose the definition
\begin{eqnarray}
\widehat{R}_{k}^{(N,a,{\rm B})}(z)&:=&\frac{1}{Z_{0|0}^{(N,a,{\rm B})}}\frac{1}{N!}\int_{\mathbb R_+^{N}}d[z']\frac{ \Delta_N^2(z') \prod_{j=1}^N {z'_j}^a\e^{-z'_j}}{\prod_{1 \le i < j \le N}(z'_i + z'_j)}\prod_{j=1}^k\left(\frac{1}{N}\sum_{i=1}^N\delta(z_j-z'_i)\right)\nonumber\\
  &=&\frac{1}{Z_{0|0}^{(N,a,{\rm B})}}\lim_{\varepsilon\to0}\sum_{L_j=\pm}\prod_{j=1}^k\left(\frac{L_j}{2\pi\imath N}\frac{\partial}{\partial \widetilde{z}_j}\right)Z_{k|k}^{(N,a,{\rm B})}(\widetilde{z}+\imath L\varepsilon,z)\biggl|_{\widetilde{z}=z},\nonumber\\
 \label{correl-Bures-def}
\end{eqnarray}
where $x=(x_1,\ldots,x_k)$, and $L=(L_1,\ldots,L_k)$, including the ``self-energy'' terms. From this quantity we can easily read off the correlation function without the ``self-energy'' terms
\begin{eqnarray}
R_{k}^{(N,a,{\rm B})}(z)&:=&\frac{1}{Z_{0|0}^{(N,a,{\rm B})}}\frac{1}{N!}\int_{\mathbb R_+^{N-k}}\prod_{j=k+1}^Ndz_j\frac{ \Delta_N^2(z') \prod_{j=1}^N {z'_j}^a\e^{-z'_j}}{\prod_{1 \le i < j \le N}(z'_i + z'_j)}.\label{conn-correl-Bures-def}
\end{eqnarray}
Also these two eigenvalue correlation functions are related via
\begin{equation}\label{connection-Bures-def}
\widehat{R}_{k}^{(N,a,{\rm B})}(z)=\frac{N!}{(N-k)!N^{k}}R_{k}^{(N,a,{\rm B})}(z)+{\rm lower\ order\ terms},
\end{equation}
where the lower order terms comprise the correlation functions $R_{k-1}^{(N,a,{\rm B})},$\\ $R_{k-2}^{(N,a,{\rm B})},\ldots$ which are all proportional to some Dirac delta-functions like $\delta(z_i-z_j)$ with $i\neq j$.

Our second aim is to establish a relation between the Bures and the Cauchy two-matrix model. Thereby we show in subsection~\ref{subsec:BurestoCauchy} that each square of the partition function~\eqref{part-Bures-def} for the Bures measure can be expressed as a partition function~\eqref{part-Cauchy-def} for the Cauchy two-matrix model. However to make sense of this relation it has to be inverted. This means we have to take the square root correctly such that we do not lose any algebraical structure which we had for the Cauchy two-matrix ensemble. For the Cauchy two-matrix ensemble we have recalled in subsection~\ref{subsec:Partition} that it  corresponds to a determinantal point process. In subsection~\ref{subsec:CauchytoBures} we show that this determinantal point process carries over to a Pfaffian point process for the Bures ensemble. In this way we calculate the skew-orthogonal polynomials, the kernels for the partition function~\eqref{part-Bures-def} (both in subsection~\ref{subsec:CauchytoBures}), and the kernels for the $k$-point correlation function~\eqref{conn-correl-Bures-def} without ``self-energy'' terms (in subsection~\ref{subsec:Correlation-Bures}).

\subsection{Going from Bures to Cauchy}\label{subsec:BurestoCauchy}

Some statements we make in subsections~\ref{subsec:BurestoCauchy} and \ref{subsec:CauchytoBures} can be applied to more general weights than the one of the joint probability density~\eqref{1.1}. Therefore let us define the partition functions
\begin{eqnarray}
Z^{(N,{\rm C})}_{k_1|l_1;k_2|l_2}[\alpha](\kappa_1,\lambda_1;\kappa_2,\lambda_2)&:=&\frac{1}{(N!)^2}\int_{\mathbb R_+^{2N} } d[x]d[y] {
  \Delta_N^2(x)  \Delta_N^2(y) \over \prod_{i,j=1}^{N} (x_i + y_j)}\nonumber\\
  &&\hspace*{-1cm}\times  \prod_{j=1}^{N} \left( \alpha(x_j) \alpha(y_j) y_j\frac{\prod_{i=1}^{l_1}(x_j-\lambda_{1,i})\prod_{i=1}^{l_2}(y_j-\lambda_{2,i})}{\prod_{i=1}^{k_1}(x_j-\kappa_{1,i})\prod_{i=1}^{k_2}(y_j-\kappa_{2,i})}\right)\nonumber\\
  \label{part-genCauchy-def}
\end{eqnarray}
for a Cauchy-like two-matrix model and
\begin{equation}
Z^{(N,{\rm B})}_{k|l}[\alpha](\kappa,\lambda):=\frac{1}{N}\int_{\mathbb R_+^{N} }d[z]\frac{ \Delta_N^2(z) }{\prod_{1 \le i < j \le N}(z_i + z_j)}\prod\limits_{j=1}^N\alpha(z_j)\frac{\prod_{i=1}^l(z_j-\lambda_i)}{\prod_{i=1}^k(z_j-\kappa_i)}\label{part-genBures-def}
\end{equation}
for a Bures-like ensemble. The weight $\alpha(z)$ is a one point weight and is in the case of the original Bures ensemble $\alpha(z)= z^a\e^{-z}$ such that we have
\begin{eqnarray}
Z^{(N,{\rm C})}_{k_1|l_1;k_2|l_2}[ z^a\e^{-z}]=Z^{(N,a,a+1{\rm C})}_{k_1|l_1;k_2|l_2}\ {\rm and}\ Z^{(N,{\rm B})}_{k|l}[ z^a\e^{-z}]=Z^{(N,a,{\rm B})}_{k|l}.
\end{eqnarray}
With these definitions we prove the following proposition in appendix~\ref{app:Prop1}.
 
 \begin{proposition}\label{Prop:Cauchy-Bures}
For an arbitrary, suitable integrable scalar function $\alpha(x)$ and two sets of variables $\lambda_1,\ldots,\lambda_l\in\mathbb{C}$ and $\kappa_1,\ldots,\kappa_k\in\mathbb{C}\setminus\mathbb{R}^+_0$ pairwise different, and $N\in\mathbb{N}$ the partition functions of the Bures-like ensemble, $Z^{(N,{\rm B})}_{k|l}[\alpha](\kappa,\lambda)$, and of the Cauchy-like two-matrix ensemble, $Z^{(N,{\rm C})}_{k|l;k|l}[\alpha](\kappa,\lambda;\kappa,\lambda)$, are related as
 \begin{eqnarray}
 \left(Z^{(N,{\rm B})}_{k|l}[\alpha](\kappa,\lambda)\right)^2&=&2^N Z^{(N,{\rm C})}_{k|l;k|l}[\alpha](\kappa,\lambda;\kappa,\lambda).\label{Prop}
  \end{eqnarray}
  \end{proposition} 
 
We remark that the derivation of Eq.~\eqref{Prop} makes use of a rewrite of the Cauchy-Vandermonde determinant 
\eqref{Ber-main} as well as the Schur Pfaffian identity ~\cite{IOTZ06}
\begin{eqnarray}\label{SP}
 \frac{\Delta_N(z)}{ \prod_{1 \le i < j \le N} (z_i + z_j)}&=&\prod_{1 \le i < j \le N} \frac{z_j - z_i}{z_j+z_i}\\
 &=& \left\{\begin{array}{cl} \displaystyle{\rm Pf} \left[ \frac{z_a-z_b}{z_a+z_b} \right]_{1\leq a,b\leq N}, & N\ {\rm even}, \\ \displaystyle{\rm Pf} \left[\begin{array}{c|c} 0 & \begin{array}{ccc} -1 & \cdots & -1 \end{array} \\ \hline \begin{array}{c} 1 \\ \vdots \\ 1 \end{array} & \displaystyle\left\{ \frac{z_a-z_b}{z_a+z_b} \right\}_{1\leq a,b\leq N} \end{array}\right], & N\ {\rm odd}. \end{array}\right.\nonumber
 \end{eqnarray}

As a simple corollary the normalization constants of the Bures ensemble and the Cauchy two-matrix ensemble are directly related, see Ref.~\cite{BGS08} where it was first proven.
 \begin{corollary}\label{cor:constants}
 The case $k=l=0$ of proposition~\ref{Prop:Cauchy-Bures} yields the normalization constant and explicitly reads
 \begin{eqnarray}\label{3.1}
 &&\left(\frac{1}{N!}\int_{\mathbb R_+^{N} }d[z]{ \Delta_N^2(z) \prod_{j=1}^{N} \alpha (z_j) \over \prod_{1 \le i < j \le N}(z_i + z_j)}
 \right)^2 \nonumber \\
 &=& \frac{2^N}{(N!)^2}
  \int_{\mathbb R_+^{2N} }d[x]d[y]{\prod_{j=1}^{N}  \alpha(x_j) \alpha(y_j) y_j \,
  \Delta_N^2(x)  \Delta_N^2(y) \over \prod_{i,j=1}^{N} (x_i + y_j)}.
  \end{eqnarray}
  \end{corollary}

Our proportionality constants in~\eqref{3.1} are different to those in Ref.~\cite{BGS08}. In fact a check can be made on this latter point, by making the choice $\alpha(x) = x^a \e^{-x}$. The LHS of  \eqref{3.1} can then be evaluated using matrix integral methods~\cite{ZS03},
 \begin{eqnarray}
 Z^{(N,a,{\rm B})}_{0|0}&=&\frac{1}{N!}\int_{\mathbb{R}_+^N} d[z] \prod_{j = 1}^N z_j^{a} \e^{-z_j} \prod_{1 \le i < j \le N}
{(z_j - z_i)^2 \over z_j + z_i} \nonumber \\
&=& \pi^{N/2} 2^{-N^2 - 2Na}
\prod_{j=0}^{N-1} {\Gamma(1 + j) \Gamma(2a + 2 + j) \over \Gamma(j+a+3/2)}.\label{SZn}
\end{eqnarray}
while from \cite[Eq.~(2-6)]{BGS08} (note that Eq.~(2-7) of \cite{BGS08}, obtained from Eq.~(2-6) using the duplication formula
for the Gamma function, contains a typo) we deduce that
\begin{eqnarray}\label{SZn1}
 Z^{(N,a,a+1,{\rm C})}_{0|0;0|0}&=&\frac{1}{(N!)^2}\int_{\mathbb R_+^{2N}}d[x]d[y]{\prod_{j=1}^{N}x_j^a \e^{-x_j}  y_j^{a+1} \e^{-y_j}  \,
  \Delta^2(x)  \Delta^2(y) \over \prod_{i,j=1}^{N} (x_i + y_j)}  \nonumber \\
  &=&\prod_{j=0}^{N-1} \Big ( {\Gamma(1+j) \Gamma(2a + 2 + j) \over \Gamma(j+a+3/2)} \Big )^2
  { \pi \over 2^{4j+ 4a + 3}},
  \end{eqnarray}
cf.~Eq.~\eqref{normalization-Cauchy}. Using these in Eq.~\eqref{3.1} the claimed proportionality of the identity is verified.
 
\subsection{Going from Cauchy to Bures}\label{subsec:CauchytoBures}
 
Looking at the established relationship between the Bures ensemble and the Cauchy two matrix model one can ask if one can invert the result of Proposition~\ref{Prop:Cauchy-Bures} and derive all correlation functions for the Bures ensemble from the Cauchy two-matrix model. Indeed one can readily take the square root of the partition functions $Z^{(N,a,{\rm C})}_{k|l;k|l}$ to find $Z^{(N,a,{\rm B})}_{k|l}$. However when doing so we may loose the algebraic structure which is a determinantal point process for the Cauchy two-matrix ensemble, see Eq.~\eqref{Cauchy-det-identity} for $k_1=k_2=k$ and $l_1=l_2=l$. 
We would expect a Pfaffian expression for the Bures ensemble when taking the square root of Eq.~\eqref{Cauchy-det-identity}. Indeed this is the case due to the Schur Pfaffian identity~\eqref{SP}. Then the joint probability density has the form of the class of ensembles discussed in \cite{KG10b} with the two-point weight $g(z_1,z_2)=z_1^az_2^a e^{-z_1-z_2}(z_1-z_2)/(z_1+z_2)$. Therefore the partition function~\eqref{part-genBures-def} has the representation ($\widetilde{N}=N+l-k>1$)
\begin{eqnarray}
&&Z^{(N,{\rm B})}_{k|l}[\alpha](\kappa,\lambda)=(-1)^{k(k-1)/2+l(l-1)/2}\frac{Z^{(\widetilde{N},{\rm B})}_{0|0}[\alpha]}{B_{k|l}(\kappa;\lambda)}\nonumber\\
&&\times\Pf\left[\begin{array}{c|c} \displaystyle (\kappa_i-\kappa_j)\frac{Z^{(\widetilde{N}+2,{\rm B})}_{2|0}[\alpha](\kappa_i,\kappa_j)}{Z^{(\widetilde{N},{\rm B})}_{0|0}[\alpha]} & \displaystyle \frac{1}{\kappa_i-\lambda_j}\frac{Z^{(\widetilde{N},{\rm B})}_{1|1}[\alpha](\kappa_i,\lambda_j)}{Z^{(\widetilde{N},{\rm B})}_{0|0}[\alpha]} \\ \hline \displaystyle \frac{1}{\lambda_i-\kappa_j}\frac{Z^{(\widetilde{N},{\rm B})}_{1|1}[\alpha](\kappa_i,\lambda_j)}{Z^{(\widetilde{N},{\rm B})}_{0|0}[\alpha]} & \displaystyle (\lambda_i-\lambda_j)\frac{Z^{(\widetilde{N}-2,{\rm B})}_{0|2}[\alpha](\lambda_i,\lambda_j)}{Z^{(\widetilde{N},{\rm B})}_{0|0}[\alpha]} \end{array}\right]\nonumber\\
\label{Pfaff-struc-Bures-even}
\end{eqnarray}
for $k+l$ even and
\begin{eqnarray}
&&Z^{(N,{\rm B})}_{k|l}[\alpha](\kappa,\lambda)=(-1)^{k(k-1)/2+l(l-1)/2}\frac{Z^{(\widetilde{N}+1,{\rm B})}_{0|0}[\alpha]}{B_{k|l}(\kappa;\lambda)}\nonumber\\
&&\hspace*{-1cm}\times\Pf\left[\begin{array}{c|c|c} \displaystyle (\kappa_i-\kappa_j)\frac{Z^{(\widetilde{N}+3,{\rm B})}_{2|0}[\alpha](\kappa_i,\kappa_j)}{Z^{(\widetilde{N}+1,{\rm B})}_{0|0}[\alpha]} & \displaystyle \frac{1}{\kappa_i-\lambda_j}\frac{Z^{(\widetilde{N}+1,{\rm B})}_{1|1}[\alpha](\kappa_i,\lambda_j)}{Z^{(\widetilde{N}+1,{\rm B})}_{0|0}[\alpha]} & \displaystyle\frac{Z^{(\widetilde{N}+1,{\rm B})}_{1|0}[\alpha](\kappa_i)}{Z^{(\widetilde{N}+1,{\rm B})}_{0|0}[\alpha]} \\ \hline \displaystyle \frac{1}{\lambda_i-\kappa_j}\frac{Z^{(\widetilde{N}+1,{\rm B})}_{1|1}[\alpha](\kappa_i,\lambda_j)}{Z^{(\widetilde{N}+1,{\rm B})}_{0|0}[\alpha]}  & \displaystyle (\lambda_i-\lambda_j)\frac{Z^{(\widetilde{N}-1,{\rm B})}_{0|2}[\alpha](\lambda_i,\lambda_j)}{Z^{(\widetilde{N}+1,{\rm B})}_{0|0}[\alpha]} & \displaystyle\frac{Z^{(\widetilde{N}-1,{\rm B})}_{0|1}[\alpha](\lambda_i)}{Z^{(\widetilde{N}+1,{\rm B})}_{0|0}[\alpha]} \\ \hline \displaystyle-\frac{Z^{(\widetilde{N}+1,{\rm B})}_{1|0}[\alpha](\kappa_j)}{Z^{(\widetilde{N}+1,{\rm B})}_{0|0}[\alpha]} & \displaystyle-\frac{Z^{(\widetilde{N}-1,{\rm B})}_{0|1}[\alpha](\lambda_j)}{Z^{(\widetilde{N}+1,{\rm B})}_{0|0}[\alpha]}  & 0 \end{array}\right]\nonumber\\
\label{Pfaff-struc-Bures-odd}
\end{eqnarray}
for $k+l$ odd. The indices take the values $1\leq i,j\leq k$ in the first few rows and columns and $1\leq i,j\leq l$ in the second set of rows and columns in both equations. See Ref.~\cite{Ra00}, Appendix~C of Ref.~\cite{KG10} 
or \cite[\S 6.3.2\&\S6.3.3]{Fo10} for a general calculation of integrals over a product of a Pfaffian and a determinant. After having this Pfaffian the kernels can be identified by choosing particular values of $k$ and $l$.

The results~\eqref{Pfaff-struc-Bures-even} and \eqref{Pfaff-struc-Bures-odd} cannot be obtained in a trivial way by only taking the square root of the partition functions of the Cauchy two-matrix model since the determinant~\eqref{Cauchy-det-identity} is not over an anti-symmetric matrix and the fact that it is an exact square is obscured. To uncover this fact one needs relations between the two-point partition functions~(\ref{two-point-Cauchy-a}-\ref{two-point-Cauchy-d}) of the Cauchy two-matrix model. We underline that those relations as well as the determinantal structure with its kernels in terms of these two-point partition functions also hold for general weight $\alpha$.  We derive these relations in the proof given in appendix~\ref{app:Prop2}, of the following proposition.

\begin{proposition}\label{Prop:det-structure-antisym}
With the requirements of proposition~\ref{Prop:Cauchy-Bures} and $N+l-k>1$ the partition function of the Cauchy-like two-matrix model can be rewritten as
\begin{eqnarray}
&&Z^{(N,{\rm C})}_{k|l;k|l}[\alpha](\kappa,\lambda;\kappa,\lambda)\nonumber\\
&=&\frac{Z^{(N+l-k,{\rm C})}_{0|0;0|0}[\alpha]}{{\rm B}_{k|l}^2(\kappa;\lambda)}\det\left[\begin{array}{c|c} \widehat{K}_{11}^{(N+l-k+1)}(\kappa_i,\kappa_j) & -\widehat{K}_{01}^{(N+l-k)}(\kappa_i,\lambda_j) \\ \hline \widehat{K}_{01}^{(N+l-k)}(\kappa_j,\lambda_i) & \widehat{K}_{00}^{(N+l-k-1)}(\lambda_i,\lambda_j) \end{array}\right] \label{det-antisym-even}
\end{eqnarray}
for $k+l$ even with $1\leq i,j\leq k$ in the first rows and columns and $1\leq i,j\leq l$ in the last ones and
\begin{eqnarray}
&&\hspace*{-1cm}Z^{(N,{\rm C})}_{k|l;k|l}[\alpha](\kappa,\lambda;\kappa,\lambda)=\frac{Z^{(N+l-k+1,{\rm C})}_{0|0;0|0}[\alpha]}{{\rm B}_{k|l}^2(\kappa;\lambda)}\nonumber\\
&&\times\det\left[\begin{array}{c|c|c}  \widehat{K}_{11}^{(N+l-k+2)}(\kappa_i,\kappa_j) & -\widehat{K}_{01}^{(N+l-k+1)}(\kappa_i,\lambda_j) & -\widehat{K}_{1}^{(N+l-k+1)}(\kappa_i) \\ \hline \widehat{K}_{01}^{(N+l-k+1)}(\kappa_j,\lambda_i) & \widehat{K}_{00}^{(N+l-k)}(\lambda_i,\lambda_j) & \widehat{K}_{0}^{(N+l-k)}(\lambda_i) \\ \hline \widehat{K}_{1}^{(N+l-k+1)}(\kappa_j) & -\widehat{K}_{0}^{(N+l-k)}(\lambda_j)  & 0 \end{array}\right]\nonumber\\ \label{det-antisym-odd}
\end{eqnarray}
for $k+l$ odd with $1\leq i,j\leq k$ in the first set of rows and columns and $1\leq i,j\leq l$ in the second set of rows and columns, cf. eq.~\eqref{det-antisym-even}. The kernels of these determinants read
\begin{eqnarray}
\widehat{K}_{11}^{(L)}(\kappa_1,\kappa_2)&=&\frac{Z^{(L,{\rm C})}_{1|0;1|0}[\alpha](\kappa_1;\kappa_2)-Z^{(L,{\rm C})}_{1|0;1|0}[\alpha](\kappa_2;\kappa_1)}{2Z^{(L-1,{\rm C})}_{0|0;0|0}[\alpha]},\nonumber\\
\widehat{K}_{01}^{(L)}(\kappa,\lambda)&=&\frac{Z^{(L,{\rm C})}_{0|0;1|1}[\alpha](\kappa,\lambda)+Z^{(L,{\rm C})}_{1|1;0|0}[\alpha](\kappa,\lambda)}{2 Z^{(L,{\rm C})}_{0|0;0|0}[\alpha](\kappa-\lambda)},\nonumber\\
\widehat{K}_{00}^{(L)}(\lambda_1,\lambda_2)&=&\frac{Z^{(L,{\rm C})}_{0|1;0|1}[\alpha](\lambda_2;\lambda_1)-Z^{(L,{\rm C})}_{0|1;0|1}[\alpha](\lambda_1;\lambda_2)}{2Z^{(L+1,{\rm C})}_{0|0;0|0}[\alpha]},\nonumber\\
\widehat{K}_{1}^{(L)}(\kappa)&=&-\frac{Z^{(L,{\rm C})}_{0|0;1|0}[\alpha](\kappa)+Z^{(L,{\rm C})}_{1|0;0|0}[\alpha](\kappa)}{2 Z^{(L,{\rm C})}_{0|0;0|0}[\alpha]},\nonumber\\
\widehat{K}_{0}^{(L)}(\lambda)&=&\frac{Z^{(L,{\rm C})}_{0|0;0|1}[\alpha](\lambda)-Z^{(L,{\rm C})}_{0|1;0|0}[\alpha](\lambda)}{2Z^{(L+1,{\rm C})}_{0|0;0|0}[\alpha]}\label{Kernels-Prop:det-antisym}
\end{eqnarray}
for $L\in\mathbb{N}$ and $L>1$.
\end{proposition}
 
Now we can take the square root and find the following corollary and one of our main results.
\begin{corollary}\label{Pfaffian-Cauchy-Bures}
 With the requirements of proposition~\ref{Prop:Cauchy-Bures} the partition function of the Bures-like ensemble can be expressed in terms of two- and one-point partition functions of the Cauchy-like two-matrix ensemble  according to
\begin{eqnarray}\label{Pf-antisym-even}
\hspace*{-1cm}Z^{(N,{\rm B})}_{k|l}[\alpha](\kappa,\lambda)&=&(-1)^{k(k-1)/2+l(l-1)/2}2^{N/2}\frac{\sqrt{Z^{(N+l-k,{\rm C})}_{0|0;0|0}[\alpha]}}{{\rm B}_{k|l}(\kappa;\lambda)}\\
&&\times\Pf\left[\begin{array}{c|c} -s\widehat{K}_{11}^{(N+l-k+1)}(\kappa_i,\kappa_j) & -\widehat{K}_{01}^{(N+l-k)}(\kappa_i,\lambda_j) \\ \hline \widehat{K}_{01}^{(N+l-k)}(\kappa_j,\lambda_i) & -s\widehat{K}_{00}^{(N+l-k-1)}(\lambda_i,\lambda_j) \end{array}\right]\nonumber
\end{eqnarray}
for $k+l$ even and
\begin{eqnarray}\label{Pf-antisym-odd}
&&\hspace*{-1cm}Z^{(N,{\rm B})}_{k|l}[\alpha](\kappa,\lambda)=(-1)^{k(k-1)/2+l(l-1)/2}2^{N/2}\frac{\sqrt{Z^{(N+l-k+1,{\rm C})}_{0|0;0|0}[\alpha]}}{{\rm B}_{k|l}(\kappa;\lambda)}\\
&&\hspace*{0cm}\times\Pf\left[\begin{array}{c|c|c}  -s\widehat{K}_{11}^{(N+l-k+2)}(\kappa_i,\kappa_j) & -\widehat{K}_{01}^{(N+l-k+1)}(\kappa_i,\lambda_j) & -\widehat{K}_{1}^{(N+l-k+1)}(\kappa_i) \\ \hline \widehat{K}_{01}^{(N+l-k+1)}(\kappa_j,\lambda_i) & -s\widehat{K}_{00}^{(N+l-k)}(\lambda_i,\lambda_j) & -s\widehat{K}_{0}^{(N+l-k)}(\lambda_i) \\ \hline \widehat{K}_{1}^{(N+l-k+1)}(\kappa_j) & s\widehat{K}_{0}^{(N+l-k)}(\lambda_j)  & 0 \end{array}\right]\nonumber
\end{eqnarray}
for $k+l$ odd. The indices $i$ and $j$ take the same values as in Eqs.~\eqref{det-antisym-even} and \eqref{det-antisym-odd}, respectively. The variable $s$ is the sign of the mean value of the difference of the two variable sets of the Cauchy-like two-matrix ensemble
\begin{eqnarray}
\widehat{Z}^{(L,{\rm C})}[\alpha]=\frac{1}{(L!)^2}\int_{\mathbb{R}_+^{2L}}d[x]d[y]\frac{\Delta_{L}^2(x)\Delta_{L}^2(y)}{\prod_{i,j=1}^{L}(x_i+y_j)}\left(\prod\limits_{j=1}^{L}\alpha(x_j)y_j\alpha(y_j)\right)\sum_{j=1}^L\left(x_j-y_j\right)\nonumber\\ \label{mean-diff}
\end{eqnarray}
and
\begin{equation}\label{sign-def}
 s={\rm sign}\, \widehat{Z}^{(L,{\rm C})}[\alpha].
\end{equation}
Additionally, in Eqs.~\eqref{Pf-antisym-even} and \eqref{Pf-antisym-odd} the kernels can be identified with partition functions of the Bures-like ensemble as well as with partition functions of the Cauchy-like two-matrix ensemble, 
\begin{eqnarray}
\frac{Z^{(N,{\rm B})}_{2|0}[\alpha](\kappa_1,\kappa_2)}{Z^{(N,{\rm B})}_{0|0}[\alpha]}&=& \frac{1}{\kappa_2-\kappa_1}\frac{Z^{(N-1,{\rm C})}_{1|0;1|0}[\alpha](\kappa_1;\kappa_2)-Z^{(N-1,{\rm C})}_{1|0;1|0}[\alpha](\kappa_2;\kappa_1)}{\widehat{Z}^{(N-1,{\rm C})}[\alpha]},\nonumber\\
\frac{Z^{(N,{\rm B})}_{1|1}[\alpha](\kappa,\lambda)}{Z^{(N,{\rm B})}_{0|0}[\alpha]}&=& \frac{Z^{(N,{\rm C})}_{0|0;1|1}[\alpha](\kappa,\lambda)+Z^{(N,{\rm C})}_{1|1;0|0}[\alpha](\kappa,\lambda)}{2 Z^{(N,{\rm C})}_{0|0;0|0}[\alpha]},\nonumber\\
\frac{Z^{(N,{\rm B})}_{0|2}[\alpha](\lambda_1,\lambda_2)}{Z^{(N,{\rm B})}_{0|0}[\alpha]}&=& \frac{1}{\lambda_1-\lambda_2}\frac{Z^{(N+1,{\rm C})}_{0|1;0|1}[\alpha](\lambda_1;\lambda_2)-Z^{(N+1,{\rm C})}_{0|1;0|1}[\alpha](\lambda_2;\lambda_1)}{\widehat{Z}^{(N+1,{\rm C})}[\alpha]},\nonumber\\
\frac{Z^{(N,{\rm B})}_{1|0}[\alpha](\kappa)}{Z^{(N,{\rm B})}_{0|0}[\alpha]}&=& \frac{Z^{(N,{\rm C})}_{0|0;1|0}[\alpha](\kappa)+Z^{(N,{\rm C})}_{1|0;0|0}[\alpha](\kappa)}{2Z^{(N,{\rm C})}_{0|0;0|0}[\alpha]},\nonumber\\
\frac{Z^{(N,{\rm B})}_{0|1}[\alpha](\lambda)}{Z^{(N,{\rm B})}_{0|0}[\alpha]}&=& \frac{Z^{(N+1,{\rm C})}_{0|1;0|0}[\alpha](\lambda)-Z^{(N+1,{\rm C})}_{0|0;0|1}[\alpha](\lambda)}{\widehat{Z}^{(N+1,{\rm C})}[\alpha]}.\nonumber\\ \label{kernels-Bures}
\end{eqnarray}
\end{corollary} 

We emphasize that the normalization in the relations~\eqref{kernels-Bures} agree with the formerly chosen one but in the way given in Eq.~\eqref{kernels-Bures} they can be easily checked. The overall sign can be identified with the one in Eqs.~\eqref{Pfaff-struc-Bures-even} and \eqref{Pfaff-struc-Bures-odd}. Moreover we underline that corollary~\ref{Pfaffian-Cauchy-Bures} provides only a way to derive all spectral correlations of the Bures measure with the help of the Cauchy two-matrix model. In contrast, proposition~\ref{Prop:Cauchy-Bures} does not imply that the correlations of all Cauchy two-matrix models are determined by the Bures measure since the weights for the two sets $\{x_i\}$ and $\{y_j\}$ have to be the same up to a factor $y_j$.

Thus let us come back to the original problem where $\alpha(z)=z^a \e^{-z}$. For this measure we know already the normalization constants $ Z^{(L,a,{\rm B})}_{0|0}$ and $Z^{(L,a,a+1,{\rm C})}_{0|0;0|0}$, see Eqs.~\eqref{SZn} and \eqref{SZn1}, respectively. The third normalization constant appearing in Eq.~\eqref{kernels-Bures} is
\begin{eqnarray}
 \widehat{Z}^{(L,a,a+1,{\rm C})}&=&\int_{0}^{2\pi}\frac{d\varphi}{2\pi}\e^{-\imath(2n-1)\varphi}(\widetilde{p}_{2n}^{(a,a+1)}(\e^{\imath\varphi})-p_{2n}^{(a,a+1)}(\e^{\imath\varphi}))\nonumber\\
 &=&-L\frac{2a+L+1}{2a+2L+1}Z^{(L,a,a+1,{\rm C})}_{0|0;0|0},\label{normalization-B-C}
\end{eqnarray}
which is essentially the $(2L-1)$st coefficient of the difference of  the two bi-orthogonal polynomials of order $2n$ for the Cauchy two-matrix model, see\\ Eqs.~\eqref{polynomials-Cauchy-a} and \eqref{polynomials-Cauchy-b}.

What can be said about the skew-orthogonal polynomials $q_n^{(a)}$ of the Bures ensemble, in particular those polynomials which are skew orthogonal with respect to the two-point weight $g^{(a)}(z_1,z_2)=(z_1z_2)^a\e^{-z_1-z_2}(z_1-z_2)/(z_1+z_2)$? First of all they have to satisfy the relations
\begin{eqnarray}
\int_{\mathbb{R}_+^2}dz_1dz_2g^{(a)}(z_1,z_2) q_{2n}^{(a)}(z_1)q_{2m}^{(a)}(z_1)&=&\int_{\mathbb{R}_+^2}dz_1dz_2g^{(a)}(z_1,z_2) q_{2n+1}^{(a)}(z_1)q_{2m+1}^{(a)}(z_1)=0,\nonumber\\
\int_{\mathbb{R}_+^2}dz_1dz_2g^{(a)}(z_1,z_2) q_{2n}^{(a)}(z_1)q_{2m+1}^{(a)}(z_1)&=&\frac{Z_{0|0}^{(2n+2,a,{\rm B})}}{Z_{0|0}^{(2n,a,{\rm B})}}\delta_{mn}\nonumber\\
\hspace*{-0.3cm}&&\hspace*{-2.7cm}=\frac{\pi}{16^{2n+a+1}}\frac{(2n+1)!(2n)!\Gamma(2n+2a+3)\Gamma(2n+2a+2)}{\Gamma(2n+a+5/2)\Gamma(2n+a+3/2)}\delta_{mn}\nonumber\\ \label{orth-rel-Bures}
\end{eqnarray}
for all $m,n\in\mathbb{N}$. For the polynomials of even order it is well known~\cite{Eynard,AKP10} that it is simply the average of one characteristic polynomial. It takes the following form shown in different representations, i.e. in terms of a partition function, the bi-orthogonal polynomials~\eqref{polynomials-Cauchy-a} and \eqref{polynomials-Cauchy-b}, a finite explicit sum,  a generalized hypergeometric function, and a Meijer G-function, respectively,
\begin{eqnarray}
 q_{2n}^{(a)}(x)&=&\frac{ Z^{(2n,a,{\rm B})}_{0|1}(x)}{ Z^{(2n,a,{\rm B})}_{0|0}}\nonumber\\
 &=&\lim_{y\to\infty}\frac{y^{2n}}{p_{2n+1}^{(a,b)}(y)-\widetilde{p}_{2n+1}^{(a,b)}(y)}\left(p_{2n+1}^{(a,b)}(x)-\widetilde{p}_{2n+1}^{(a,b)}(x)\right)\nonumber\\
 &=&\sum_{j=0}^{2n} (-1)^{j}\left(\begin{array}{c} 2n \\ j \end{array}\right)\frac{\Gamma(2a+2n+j+3)\Gamma(2a+2n+2)\Gamma(a+2n+2)}{\Gamma(2a+4n+3)\Gamma(2a+j+2)\Gamma(a+j+2)}x^j\nonumber\\
 &=& \frac{(2a+2)_{2n}(a+2)_{2n}}{(2a+2n+3)_{2n}}\, _2F_2\left(\left. -2n,2a+2n+3 \atop a+2,2a+2\right|x\right)\nonumber\\
 &=&\frac{(2n)!\Gamma(2a+2n+2)\Gamma(a+2n+2)}{\Gamma(2a+4n+3)}G_{2,3}^{1,1} \left( \left. {-2a-2n-2;2n+1 \atop 0;-2a-1,-a-1} \right|x\right).\nonumber\\
 \label{polynomials-Bures-even}
\end{eqnarray}
It is also well-known~\cite{AKP10} what the odd polynomials look like in terms of partition functions, namely
\begin{eqnarray}
 q_{2n+1}^{(a)}(x)=\frac{1}{ Z^{(2n,a,{\rm B})}_{0|0}}\frac{1}{(2n)!}\int_{\mathbb{R}_+^{2n}}d[z]\frac{\Delta_{2n}^2(z)\prod_{j=1}^{2n}z_j^a(x-z_j)\e^{-z_j}}{\prod_{1\leq i<j\leq 2n}(z_i+z_j)}\left(x+\sum_{j=1}^{2n}z_j+c\right),\nonumber\\
 \label{polynomials-Bures-odd-a}
\end{eqnarray}
where $c$ is an arbitrary constant and reflects the ambiguity of the solution of the skew-orthogonality relations~\eqref{orth-rel-Bures}. We make use of fixing this constant later on to simply our results.

The constant $c$ as well as the variable $x$ can be pulled out the integral~\eqref{polynomials-Bures-odd-a} leaving the polynomial $q_{2n}^{(a)}$ as a factor. The trace can be written as a derivative of an auxiliary parameter $t$ which is introduced in the exponent, in particular we replace the two point weight $g^{(a)}(z_1,z_2)\to g_t^{(a)}(z_1,z_2)=(z_1z_2)^a\e^{-t(z_1+z_2)}(z_1-z_2)/(z_1+z_2)$. Then the integration variables $z$ can be rescaled, $z\to z/t$ such that we find the identity
\begin{equation}
 q_{2n+1}^{(a)}(x)=(x+c)q_{2n}^{(a)}(x)-\left.\frac{\partial}{\partial t}t^{-n(2n+3+2a)}q_{2n}^{(a)}(tx)\right|_{t=1}.
 \label{polynomials-Bures-odd-b}
\end{equation}
Fixing now the constant $c=-n(2n+3+2a)$ we find the simple result
\begin{eqnarray}
 q_{2n+1}^{(a)}(x)&=&x\left(1-\frac{\partial}{\partial x}\right)q_{2n}^{(a)}(x)\nonumber\\
 &=&\frac{(2n)!\Gamma(2a+2n+2)\Gamma(a+2n+2)}{\Gamma(2a+4n+3)}\left(G_{2,3}^{1,1} \left( \left. {-2a-2n-1;2n+2 \atop 1;-2a,-a} \right|x\right)\right.\nonumber\\
 &&\left.-G_{3,4}^{1,2} \left( \left. {0,-2a-2n-2;2n+1 \atop 0;1,-2a-1,-a-1} \right|x\right)\right).\nonumber\\
 \label{polynomials-Bures-odd-c}
\end{eqnarray}
Note that we do not need the polynomials of odd order in our approach. We only show them for the sake of completeness.

Also the kernels of the Pfaffians~\eqref{Pfaff-struc-Bures-even} and \eqref{Pfaff-struc-Bures-odd} can be read off yielding the following corollary.
\begin{corollary}\label{cor:Pfaffian-Bures}
 With the requirements of proposition~\ref{Prop:Cauchy-Bures} and $\widetilde{N}=N+l-k$ the kernels of the Pfaffian representations~\eqref{Pfaff-struc-Bures-even} and \eqref{Pfaff-struc-Bures-odd} for the partition function~\eqref{part-Bures-def} of the original Bures ensemble ($\alpha(z)=z^a\e^{-z}$) are
\begin{eqnarray}
&&(\lambda_1-\lambda_2)\frac{Z^{(\widetilde{N}-2,a,{\rm B})}_{0|2}(\lambda_1,\lambda_2)}{Z^{(\widetilde{N},a,{\rm B})}_{0|0}}\nonumber\\
&=&-\frac{1}{4}\int_0^1dt\biggl[G_{2,3}^{1,1} \left( \left. {-2a-\widetilde{N}-1;\widetilde{N} \atop 0;-a,-2a-1} \right|t\lambda_1\right)G_{2,3}^{1,1} \left( \left. {-2a-\widetilde{N}-1;\widetilde{N} \atop 0;-a-1,-2a-1} \right|t\lambda_2\right)\nonumber\\
&&-G_{2,3}^{1,1} \left( \left. {-2a-\widetilde{N}-1;\widetilde{N} \atop 0;-a,-2a-1} \right|t\lambda_2\right)G_{2,3}^{1,1} \left( \left. {-2a-\widetilde{N}-1;\widetilde{N} \atop 0;-a-1,-2a-1} \right|t\lambda_1\right)\biggl],\nonumber\\
 \label{kernels-Bures-original-a}
\end{eqnarray}
\begin{eqnarray}
&&\frac{1}{\kappa-\lambda}\frac{Z^{(\widetilde{N},a,{\rm B})}_{1|1}(\kappa,\lambda)}{Z^{(\widetilde{N},a,{\rm B})}_{0|0}}=\frac{1}{\kappa-\lambda}\nonumber\\
&&+\frac{1}{2}\int_0^1dt\biggl[G_{2,3}^{1,1} \left( \left. {-2a-\widetilde{N}-1;\widetilde{N} \atop 0;-a,-2a-1} \right|t\lambda\right)G_{2,3}^{3,1} \left( \left. {-\widetilde{N};\widetilde{N}+2a+1 \atop 0,a,2a+1} \right|-t\kappa\right)\nonumber\\
&&+G_{2,3}^{1,1} \left( \left. {-2a-\widetilde{N}-1;\widetilde{N} \atop 0;-a-1,-2a-1} \right|t\lambda\right)G_{2,3}^{3,1} \left( \left. {-\widetilde{N};\widetilde{N}+2a+1 \atop 0,a+1,2a+1} \right|-t\kappa\right)\biggl],\nonumber\\
 \label{kernels-Bures-original-b}
\end{eqnarray}
\begin{eqnarray}
&&(\kappa_1-\kappa_2)\frac{Z^{(\widetilde{N}+2,a,{\rm B})}_{2|0}(\kappa_1,\kappa_2)}{Z^{(\widetilde{N},a,{\rm B})}_{0|0}}=\int_{\mathbb{R}_+^2}dxdy\frac{(xy)^a(y-x)\e^{-x-y}}{(\kappa_1-x)(\kappa_2-y)(x+y)}\nonumber\\
&&\hspace*{-1cm}+(\kappa_1\kappa_2)^a\int_0^1dt\biggl[\kappa_2G_{2,3}^{3,1} \left( \left. {-\widetilde{N}-a;\widetilde{N}+a+1 \atop 0,-a,a+1} \right|-t\kappa_1\right)G_{2,3}^{3,1} \left( \left. {-\widetilde{N}-a-1;\widetilde{N}+a \atop 0,-a-1,a} \right|-t\kappa_2\right)\nonumber\\
&&\hspace*{-1cm}-\kappa_1G_{2,3}^{3,1} \left( \left. {-\widetilde{N}-a;\widetilde{N}+a+1 \atop 0,-a,a+1} \right|-t\kappa_2\right)G_{2,3}^{3,1} \left( \left. {-\widetilde{N}-a-1;\widetilde{N}+a \atop 0,-a-1,a} \right|-t\kappa_1\right)\nonumber\\
&&\hspace*{-1cm}-\kappa_2G_{2,3}^{3,1} \left( \left. {-a;a+1 \atop 0,-a,a+1} \right|-t\kappa_1\right)G_{2,3}^{3,1} \left( \left. {-a-1;a \atop 0,-a-1,a} \right|-t\kappa_2\right)\nonumber\\
&&\hspace*{-1cm}+\kappa_1G_{2,3}^{3,1} \left( \left. {-a;a+1 \atop 0,-a,a+1} \right|-t\kappa_2\right)G_{2,3}^{3,1} \left( \left. {-a-1;a \atop 0,-a-1,a} \right|-t\kappa_1\right)\biggl]
 \label{kernels-Bures-original-c}
\end{eqnarray}
for the two-point kernels and
\begin{eqnarray}
\frac{Z^{(\widetilde{N}-1,a,{\rm B})}_{0|1}(\lambda)}{Z^{(\widetilde{N}+1,a,{\rm B})}_{0|0}}=\frac{2^{2(\widetilde{N}+a)}\Gamma[\widetilde{N}+a+3/2]}{\widetilde{N}!\Gamma[\widetilde{N}+2a+2]}G_{2,3}^{1,1} \left( \left. {-\widetilde{N}-2a-1;\widetilde{N} \atop 0;-2a-1,-a-1} \right|\lambda\right),\nonumber\\
 \label{kernels-Bures-original-d}
\end{eqnarray}
and
\begin{eqnarray}
\frac{Z^{(\widetilde{N}+1,a,{\rm B})}_{1|0}(\kappa)}{Z^{(\widetilde{N}+1,a,{\rm B})}_{0|0}}=-\frac{1}{\widetilde{N}!\Gamma[\widetilde{N}+a+1]}G_{2,3}^{3,1} \left( \left. {-\widetilde{N}-1;\widetilde{N}+2a \atop -1,a-1,2a} \right|-\kappa\right)\nonumber\\
 \label{kernels-Bures-original-e}
\end{eqnarray}
for the partition functions with only one characteristic polynomial.
\end{corollary} 
With the help of this corollary we are now ready to find all $k$-point correlation functions of the Bures ensemble.
 
\subsection{Correlation functions of the Bures ensemble}\label{subsec:Correlation-Bures}
 
Proposition~\ref{Prop:Cauchy-Bures} relates the partition functions of the Bures ensemble with those of the Cauchy two-matrix model. Thereby the individual level density of a random matrix has to be exchanged in the following way
\begin{equation}\label{SP1}
\sum_{j=1}^N \delta(z - z_j) \rightarrow\sum_{j=1}^N [ \delta(z - x_j) + \delta( z - y_j) \Big ]
\end{equation}
for the three species $\{x_j\}$, $\{y_j\}$, and $\{z_j\}$  in Eqs.~\eqref{part-Cauchy-def} and \eqref{part-Bures-def}, respectively. On the other hand, Eq.~\eqref{con-correl-Cauchy-result}, in particular Eqs.~\eqref{level-density-Cauchy-a} and \eqref{level-density-Cauchy-b} for the densities, gives the correlations relating to the  averaged  level densities of the two species regarded as separate entities. Thus knowledge of the eigenvalue correlations of the Cauchy two-matrix model  implies the correlations of the Bures ensemble. In particular, we have for the level densities
\begin{eqnarray}
R_{1}^{(N,a,{\rm B})}(z)&=&\lim_{\varepsilon\to0}\sum_{L=\pm1}\frac{L}{2\pi\imath N}\frac{\partial}{\partial \widetilde{z}}{\rm ln}\, Z_{1|1}^{(N,a,{\rm B})}(\widetilde{z}+\imath L\varepsilon,z)\biggl|_{\widetilde{z}=z}\nonumber\\
&=&\frac{1}{2}\lim_{\varepsilon\to0}\sum_{L=\pm1}\frac{L}{2\pi\imath N}\frac{\partial}{\partial \widetilde{z}}{\rm ln}\, Z_{1|1;1|1}^{(N,a,a+1{\rm C})}(\widetilde{z}+\imath L\varepsilon,z;\widetilde{z}+\imath L\varepsilon,z)\biggl|_{\widetilde{z}=z}\nonumber\\
&=&\frac{1}{2}\left(R_{1,0}^{(N,a,a+1,{\rm C})}(z)+R_{0,1}^{(N,a,a+1,{\rm C})}(z)\right).\label{rel-level-densities}
\end{eqnarray}
The same calculation can be done for the two-point correlation function with ``self-energy" terms,
\begin{eqnarray}
\widehat{R}_{2}^{(N,a,{\rm B})}(z_1,z_2)&=&\lim_{\varepsilon\to0}\sum_{L_1,L_2=\pm1}\frac{L_1L_2}{(2\pi\imath N)^2}\frac{\partial^2}{\partial \widetilde{z}_1\partial \widetilde{z}_2}\biggl[{\rm ln}\, Z_{2|2}^{(N,a,{\rm B})}(\widetilde{z}+\imath L\varepsilon,z)\nonumber\\
&&+{\rm ln}\, Z_{1|1}^{(N,a,{\rm B})}(\widetilde{z}_1+\imath L_1\varepsilon,z_1){\rm ln}\, Z_{1|1}^{(N,a,{\rm B})}(\widetilde{z}_2+\imath L_1\varepsilon,z_2)\biggl]\biggl|_{\widetilde{z}=z}.\nonumber\\
&&\label{two-point-densities}
\end{eqnarray}
The first term is the connected correlation function and is the analogue to the cumulant for ordinary random variables. Employing relation~\eqref{con-correl-Cauchy-result} we have
\begin{eqnarray}
&&\widehat{R}_{2}^{(N,a,{\rm B})}(z_1,z_2)\nonumber\\
&=&\frac{1}{2}\biggl[\widehat{R}_{2,0}^{(N,a,a+1,{\rm C})}(z_1,z_2)+\widehat{R}_{1,1}^{(N,a,a+1,{\rm C})}(z_1,z_2)+\widehat{R}_{1,1}^{(N,a,a+1,{\rm C})}(z_2,z_1)\nonumber\\
&&+\widehat{R}_{0,2}^{(N,a,a+1,{\rm C})}(z_1,z_2)\nonumber\\
&&-\frac{1}{2}\left(R_{1,0}^{(N,a,a+1,{\rm C})}(z_1)+R_{0,1}^{(N,a,a+1,{\rm C})}(z_1)\right)\left(R_{1,0}^{(N,a,a+1,{\rm C})}(z_2)+R_{0,1}^{(N,a,a+1,{\rm C})}(z_2)\right)\biggl]\nonumber\\
&=&\frac{1}{2}\biggl[\frac{1}{2}\left(K_{01}^{(N,a,a+1,{\rm C})}(z_1,z_1)+K_{10}^{(N,a,a+1,{\rm C})}(z_1,z_1)\right)\nonumber\\
&&\times\left(K_{01}^{(N,a,a+1,{\rm C})}(z_2,z_2)+K_{10}^{(N,a,a+1,{\rm C})}(z_2,z_2)\right)\nonumber\\
&&-K_{01}^{(N,a,a+1,{\rm C})}(z_1,z_2)K_{01}^{(N,a,a+1,{\rm C})}(z_2,z_1)-K_{10}^{(N,a,a+1,{\rm C})}(z_1,z_2)K_{10}^{(N,a,a+1,{\rm C})}(z_2,z_1)\nonumber\\
&&-K_{11}^{(N,a,a+1,{\rm C})}(z_1,z_2)K_{00}^{(N,a,a+1,{\rm C})}(z_1,z_2)-K_{11}^{(N,a,a+1,{\rm C})}(z_2,z_1)K_{00}^{(N,a,a+1,{\rm C})}(z_2,z_1)\nonumber\\
&&+\frac{1}{N}\delta(z_1-z_2)\left(K_{01}^{(N,a,a+1,{\rm C})}(z_1,z_1)+K_{10}^{(N,a,a+1,{\rm C})}(z_1,z_1)\right)\biggl].\nonumber\\
&&\label{two-point-densities-b}
\end{eqnarray}
The term proportional to the Dirac delta-function is the ``self-energy'' term and will be omitted in the following. The kernels satisfy the relations
\begin{eqnarray}
K_{00}^{(N,a,a+1,{\rm C})}(z_1;z_2)+K_{00}^{(N,a,a+1,{\rm C})}(z_1;z_2)&=&w(z_1)w(z_2),\nonumber\\
K_{01}^{(N,a,a+1,{\rm C})}(z_1,z_2)-K_{10}^{(N,a,a+1,{\rm C})}(z_1,z_2)&=&v(z_1)w(z_2),\nonumber\\
K_{11}^{(N,a,a+1,{\rm C})}(z_1;z_2)+K_{11}^{(N,a,a+1,{\rm C})}(z_1;z_2)&=&-v(z_1)v(z_2),\nonumber\\
\label{kernel-relations}
\end{eqnarray}
where the functions $w$ and $v$ can be read off from the relations satisfied by the two-point partition functions derived in appendix~\ref{app:Prop2}. With the help of these relations one can show that the two-point correlation function without the ``self-energy'' terms is equal to the Pfaffian
 \begin{eqnarray}
 R_{2}^{(N,a,{\rm B})}(z_1,z_2)&=&-\frac{N-1}{4N}\nonumber\\
 &&\times\Pf\left[\begin{array}{c|c} \Delta K_{11}^{(N,a,a+1,{\rm C})}(z_i;z_j) & \Sigma K_{01}^{(N,a,a+1,{\rm C})}(z_i;z_j) \\ \hline -\Sigma K_{01}^{(N,a,a+1,{\rm C})}(z_i;z_j) & \Delta K_{00}^{(N,a,a+1,{\rm C})}(z_j;z_i) \end{array}\right]_{1\leq i,j\leq 2}\nonumber\\
 \label{two-point-densities-c}
 \end{eqnarray}
 with the abbreviations
 \begin{eqnarray}
 \Delta K_{11}^{(N,a,a+1,{\rm C})}(z_i;z_j)&=&K_{11}^{(N,a,a+1,{\rm C})}(z_i;z_j)-K_{11}^{(N,a,a+1,{\rm C})}(z_j;z_i),\nonumber\\
 \Sigma K_{01}^{(N,a,a+1,{\rm C})}(z_i;z_j)&=&K_{01}^{(N,a,a+1,{\rm C})}(z_i,z_j)+K_{10}^{(N,a,a+1,{\rm C})}(z_i,z_j),\nonumber\\
 \Delta K_{00}^{(N,a,a+1,{\rm C})}(z_j;z_i)&=&K_{00}^{(N,a,a+1,{\rm C})}(z_j;z_i)-K_{00}^{(N,a,a+1,{\rm C})}(z_i;z_j).\label{abbre}
 \end{eqnarray}
Indeed we can also find the results~\eqref{rel-level-densities} and \eqref{two-point-densities-c} via the relation of the partition functions derived in subsection~\ref{subsec:CauchytoBures}.

Let us employ the definition~\eqref{correl-Bures-def} to the result~\eqref{Pf-antisym-even} with $k=l$. Then the only important contribution is the action of the derivatives in $\widetilde{z}$ on the prefactor $1/B_{k|k}(\widetilde{z}+\imath L\varepsilon,z)$ in front of the Pfaffian. Almost all other terms vanish under the sum of the signs $L_j=\pm1$ and in the limit $\widetilde{z}=z$ and $\varepsilon\to0$. Indeed there are also contributions from the derivatives in the diagonal entries of the off-diagonal blocks in the Pfaffian. However they yield the same kernel as the other matrix entries in the off-diagonal blocks apart from the Dirac delta-functions $\delta(z_i-z_j)$ which result from the first term $1/(\kappa-\lambda)$ of the two-point partition function~\eqref{kernels-Bures-original-b}. Omitting these Dirac delta-functions we find the $k$-point correlation function summarized in the following corollary.
\begin{corollary}\label{cor:correlation-Bures}
 Let $z_1,\ldots,z_k\in\mathbb{R}_+$ be pairwise different. Then the $k$-point correlation function without the ``self-energy'' terms  of the Bures ensemble is
 \begin{eqnarray}
 R_{k}^{(N,a,{\rm B})}(z)&=&(-1)^{k(k-1)/2}\frac{N!}{(2N)^k(N-k)!}\nonumber\\
 &&\times\Pf\left[\begin{array}{c|c} \Delta K_{11}^{(N,a,a+1,{\rm C})}(z_i;z_j) & \Sigma K_{01}^{(N,a,a+1,{\rm C})}(z_i;z_j) \\ \hline -\Sigma K_{01}^{(N,a,a+1,{\rm C})}(z_i;z_j) & \Delta K_{00}^{(N,a,a+1,{\rm C})}(z_j;z_i) \end{array}\right]_{1\leq i,j\leq k},\nonumber\\
 \label{correlation-Bures}
 \end{eqnarray}
 where we used the abbreviations~\eqref{abbre} and the kernels~\eqref{kernel-Cauchy} of the Cauchy two-matrix model.
 \end{corollary}
Thus we have traced all $k$-point correlation functions for the Bures ensemble back to the kernels of the $(k,l)$-point correlation functions~\eqref{con-correl-Cauchy-result} for the Cauchy two-matrix model. In particular the case $k=N$ yields the joint probability density  in terms of a Pfaffian. This shows that the process given by Eq.~\eqref{1.1} is a Pfaffian point process.

\section{Hard Edge Scaling Limit}\label{sec:hard}

Finally we consider the hard edge scaling limit which is the double scaling limit $N\to\infty$ with $N^2x_j$ and $a$ fixed. This limit is a direct corollary of our result in combination with \cite[Theorem 2.2]{BGS12}.

\begin{corollary}\label{cor:hard-edge}
 Let $z_1,\ldots,z_k\in\mathbb{R}_+$ be pairwise different. Then the hard edge scaling limit of the $k$-point correlation function  of the Bures ensemble is
 \begin{eqnarray}
 R_{k}^{(\infty,a,{\rm B})}(z)&=&\lim_{N\to\infty} N^{-2k}R_{k}^{(N,a,{\rm B})}\left(\frac{z}{N^2}\right)
 \label{lim-correlation-Bures}\\
 &=&\frac{(-1)^{k(k-1)/2}}{2^k}\Pf\left[\begin{array}{c|c} \Delta K_{11}^{(\infty,a)}(z_i;z_j) & \Sigma K_{01}^{(\infty,a)}(z_i;z_j) \\ \hline -\Sigma K_{01}^{(\infty,a)}(z_i;z_j) & \Delta K_{00}^{(\infty,a)}(z_j;z_i) \end{array}\right]_{1\leq i,j\leq k}.\nonumber
 \end{eqnarray}
 The kernels are given by
 \begin{eqnarray}
 \Delta K_{11}^{(\infty,a)}(z_i;z_j)&=&\lim_{N\to\infty} N^{4a}\Delta K_{11}^{(N,a,a+1,{\rm C})}\left(\frac{z_i}{N^2};\frac{z_j}{N^2}\right)\nonumber\\
 &=&z_i^az_j^a\left(\frac{z_i-z_j}{z_i+z_j}\right.\nonumber\\
 &&+\int_0^1 dt \left[z_jG_{0,3}^{2,0} \left( \left. {- \atop 0,a+1;-a} \right|tz_i\right)G_{0,3}^{2,0} \left( \left. {- \atop 0,a;-a-1} \right|tz_j\right)\right.\nonumber\\
 &&\left.\left.-z_iG_{0,3}^{2,0} \left( \left. {- \atop 0,a;-a-1} \right|tz_i\right)G_{0,3}^{2,0} \left( \left. {- \atop 0,a+1;-a} \right|tz_j\right)\right]\right),\nonumber\\
 \Sigma K_{01}^{(\infty,a)}(z_i;z_j)&=&\lim_{N\to\infty} N^{-2}\Sigma K_{01}^{(N,a,a+1,{\rm C})}\left(\frac{z_i}{N^2};\frac{z_j}{N^2}\right)\nonumber\\
 &=&\int_0^1 dt \left[G_{0,3}^{1,0} \left( \left. {- \atop 0;-a,-2a-1} \right|tz_j\right)G_{0,3}^{2,0} \left( \left. {- \atop a,2a+1;0} \right|tz_i\right)\right.\nonumber\\
 &&\hspace*{-0.5cm}\left.+G_{0,3}^{2,0} \left( \left. {- \atop a+1,2a+1;0} \right|tz_j\right)G_{0,3}^{1,0} \left( \left. {- \atop 0;-a-1,-2a-1} \right|tz_i\right)\right],\nonumber\\
 \Delta K_{00}^{(\infty,a)}(z_j;z_i)&=&\lim_{N\to\infty} N^{-4a-4}\Delta K_{00}^{(N,a,a+1,{\rm C})}\left(\frac{z_j}{N^2};\frac{z_i}{N^2}\right)\nonumber\\
 &&\hspace*{-2cm}=\int_0^1 dt t^{2a+1}\left[G_{0,3}^{1,0} \left( \left. {- \atop 0;-a,-2a-1} \right|tz_j\right)G_{0,3}^{1,0} \left( \left. {- \atop 0;-a-1,-2a-1} \right|tz_i\right)\right.\nonumber\\
 &&\hspace*{-0.5cm}\left.-G_{0,3}^{1,0} \left( \left. {- \atop 0;-a-1,-2a-1} \right|tz_j\right)G_{0,3}^{1,0} \left( \left. {- \atop 0;-a,-2a-1} \right|tz_i\right)\right].\nonumber\\ \label{ker-lim}
 \end{eqnarray}
 \end{corollary}

The proof of this corollary is straightforward and will be skipped. One has simply to combine  \cite[Theorem 2.2]{BGS12} with Eq.~\eqref{abbre}. There is also no problem with the different scalings of the three kernels due to multilinearity of the Pfaffian. Then the factors $N^{4a}$, $N^{-2}$, and $N^{-4a-4}$ can be pulled into the corresponding rows and columns and the total factor of $N$ can be counted in the overall factor.

The Meijer G-kernel in the Pfaffian is up to now unique. It would be interesting if this kernel can be found for other random matrix ensembles as well. This spectral behavior would be quite natural as long as the pair interaction of the eigenvalues near the origin, $z_i,z_j\ll1$, behave like $\exp[2{\rm ln}|z_i-z_j|-{\rm ln}|z_i+z_j|]$. Working in~\cite{FL14} tells us that this functional form gives the equilibrium problem for the Raney density indexed by $(3/2,1/2)$, the significance of this being that the evidence is that it is the Raney parameters $(p,r)$ which determine the hard edge universality class; see again~\cite{FL14}. It is to be expected that this hard edge scaling limit is independent of the confining potential $\exp[-V(z_j)]$ implying that the a condition like $\sum_j z_j=1$ does not affect this behavior because it can be rewritten in term of such a potential via a Fourier-Laplace transform, see e.g.~\cite{LZ10}. The reason why we expect this result also for a general class of ensembles is the separation of scales which is the origin of universality of spectral statistics.

\section{Discussion and Outlook}\label{sec:conclusio}

We established a relationship between the kernels of the Pfaffian point process of the Bures ensemble and the kernels of the determinantal point process of the Cauchy two-matrix model. Thereby we started from the partially known fact~\cite{BGS08} that the square of any partition function of the Bures ensemble is equal to a partition function of a Cauchy two-matrix model, see proposition~\ref{Prop:Cauchy-Bures}. Since the kernels of the Pfaffian point process can be also identified as partition functions we had  to invert this relation. Surprisingly, the square root can be made exact such that we end up with  precisely the same one-fold integrals over Meijer G-functions as were found for the Cauchy two-matrix model~\cite{BGS12}, see corollaries~\ref{cor:Pfaffian-Bures} and \ref{cor:correlation-Bures}. In particular, each of the kernels of the Bures ensemble is only a linear combination of two kernels of the Cauchy two-matrix model. Additionally the skew-orthogonal polynomials corresponding to the Bures ensemble are expressed in the bi-orthogonal polynomials of the Cauchy two-matrix model. All these relations together represent a complete exact solution of the Bures ensemble.

Two problems can now be studied. First, as already begun we can study any large $N$ asymptotics of the level statistics including the macroscopic and microscopic  level densities, the hard edge and the soft edge correlation functions, and the correlation functions in the bulk.  Here we  expect the sine-kernel for the Dyson index $\beta=2$ in the bulk of the spectrum since the level repulsion is $(\lambda_1-\lambda_2)^2$ which should be the only relevant input on the scale of the mean level spacing. Also at the soft edge we expect the standard Airy kernel behaviour for the joint probability density \eqref{1.1} and for the
fixed trace ensemble \eqref{fixed-trace} as known for the Laguerre ensemble and its fixed trace counterpart \cite{LZ10}. The latter
corresponds to the case of the Hilbert-Schmidt measure on the set of density matrices.

At the hard edge we derived that the Bures ensemble lies in a universality class which is described by Meijer G-functions and is  reminiscent to those already found for product matrices, see Refs.~\cite{AB12,ABKN14,Fo14,KS14,KZ13,NS14}. Thus the Bures ensemble is the first ensemble of such a class for which the singular values exhibit a Pfaffian point process. We expect that the Bures ensemble with the fixed trace condition~\eqref{fixed-trace} shares the same behaviour at the hard edge since the condition should only effect the upper bound on the local scale of the mean level spacing. The macroscopic level density for the Bures ensemble with an infinite rank $N-M$ fixed was already calculated in Ref.~\cite{ZS04}. However, neither the level density at finite $N$ nor when $N/(N-M)$ is fixed  were considered. The first case is important when small quantum systems as qubits and qutrits are studied~\cite{Slater04,Slater05,Slater12,schmied}. The second case with $N/(N-M)$ fixed for $N\to\infty$ is the quite natural case for any experiment which considers a prepared quantum state which is almost pure.

In the second problem, one can go back to the original Bures measure where we have to include the fixed trace condition which results in a joint probability density~\eqref{fixed-trace}.
As previously remarked, the Pfaffian based correlations found for Eq.~\eqref{1.1} now involve an auxiliary scaling parameter, and a Fourier-Laplace transform must be taken with respect to this parameter. Although there is good reason to think that this has no effect on scaling limits~\cite{LZ10}, the details of the calculation remain. The additional integral from the Fourier-Laplace transformation destroys the algebraic structure of the partition functions and the $k$-point correlation functions in terms of Pfaffians. This additional integral will have a crucial influence on the macroscopic level density and the edge behaviour at the upper bound because the support of the eigenvalues is squeezed to the allowed interval $[0,1]$.

\section*{Acknowledgements}

MK acknowledges financial support of the Alexander von Humboldt foundation and thanks the University of Melbourne for offering him an honorary appointment of two weeks within the Faculty of Science as a visitor which initiated this project. The work of PJF was supported by the Australian Research Council, grant DP140102613.

\appendix

\section{Proof of proposition~\ref{Prop:Cauchy-Bures}}\label{app:Prop1}

Let $k=l$ without loss of generality since we can take the limits $|\lambda_j|\to\infty$ or $|\kappa_j|\to\infty$ for some $j$ which reduces the partition function for $k=l$ to the general case where $k$ and $l$ are different.
  
   The proof starts with two identities, namely the Schur Pfaffian~\eqref{SP}
 and the extension
 \begin{equation}\label{enlarge}
 \Delta_N(z)\prod\limits_{j=1}^N\frac{\prod_{i=1}^k(z_j-\lambda_i)}{\prod_{i=1}^k(z_j-\kappa_i)}=\frac{{\rm B}_{k|N+k}(\kappa;z,\lambda)}{{\rm B}_{k|k}(\kappa;\lambda)}
 \end{equation}
 of the Cauchy-Vandermonde determinant~\cite{BF94,KG10}
 \begin{eqnarray}\label{Ber}
 {\rm B}_{k|N+k}(\kappa;z,\lambda)&=&\frac{ \Delta_N(z)\Delta_{k}(\kappa)\Delta_k(\lambda)}{\prod_{i,j=1}^k(\kappa_i-\lambda_j)}\prod\limits_{j=1}^N\frac{\prod_{i=1}^k(z_j-\lambda_i)}{\prod_{i=1}^k(z_j-\kappa_i)}\\
 &&\hspace*{-1cm}=(-1)^{k(k-1)/2}\det\left[\begin{array}{cc} \displaystyle\left\{z_a^{b-1}\right\}_{1\leq a,b\leq N} & \displaystyle\left\{\frac{1}{\kappa_b-z_a}\right\}\underset{1\leq b\leq k}{\underset{1\leq a\leq N}{}}\\ \displaystyle\left\{\lambda_a^{b-1}\right\}\underset{1\leq b\leq N}{\underset{1\leq a\leq k}{}}  & \displaystyle\left\{\frac{1}{\kappa_b-\lambda_a}\right\}_{1\leq a,b\leq k} \end{array}\right],  \nonumber\\
 {\rm B}_{k|k}(\kappa;\lambda)&=&\frac{\Delta_k(\kappa)\Delta_k(\lambda)}{\prod_{i,j=1}^k(\kappa_i-\lambda_j)}=(-1)^{k(k-1)/2}\det\left[\frac{1}{\kappa_a-\lambda_b}\right]_{1\leq a,b\leq k}.\nonumber
 \end{eqnarray}
which is equivalent to \eqref{Ber-main}.
 The two identities~\eqref{SP} and \eqref{enlarge} can be plugged into the partition function $Z^{(N,{\rm B})}_{k|k}[\alpha](\kappa,\lambda)$ for the Bures-like measure. Applying a modified version of de Bruijn's integral identity~\cite{Bruijn,KG10} we find
 \begin{eqnarray}\label{p1-ev}
  &&Z^{(N,{\rm B})}_{k|k}[\alpha](\kappa,\lambda)=\pm \frac{1}{{\rm B}_{k|k}(\kappa;\lambda)}{\rm Pf}\left[M_{\rm even}\right]\\
  &&\hspace*{-0.6cm}=\pm \frac{1}{{\rm B}_{k|k}(\kappa;\lambda)}{\rm Pf}\left[\begin{array}{ccc} \displaystyle\left\{M_{ab}\right\}\underset{1\leq a,b\leq N}{} & \displaystyle\left\{F_a(\kappa_b)\right\}\underset{1\leq b \leq k}{\underset{1\leq a\leq N}{}} & \displaystyle\left\{\lambda_b^{a-1}\right\}\underset{1\leq b \leq k}{\underset{1\leq a\leq N}{}} \\ \displaystyle\left\{-F_b(\kappa_a)\right\}\underset{1\leq b \leq N}{\underset{1\leq a\leq k}{}} & \displaystyle\left\{G(\kappa_a,\kappa_b)\right\}\underset{1\leq a,b\leq k}{} & \displaystyle\left\{\frac{1}{\kappa_a-\lambda_b}\right\}\underset{1\leq a,b\leq k}{} \\ \displaystyle\left\{-\lambda_a^{b-1}\right\}\underset{1\leq b \leq N}{\underset{1\leq a\leq k}{}} & \displaystyle\left\{\frac{1}{\lambda_a-\kappa_b}\right\}\underset{1\leq a,b\leq k}{} & 0 \end{array}\right]\nonumber
 \end{eqnarray}
 for even $N$ and
 \begin{eqnarray}\label{p1-odd}
 &&Z^{(N,{\rm B})}_{k|k}[\alpha](\kappa,\lambda)=\pm \frac{1}{{\rm B}_{k|k}(\kappa;\lambda)} {\rm Pf}\left[M_{\rm odd}\right]\\
  &=&\pm \frac{1}{{\rm B}_{k|k}(\kappa;\lambda)}{\rm Pf}\left[\begin{array}{c|c} 0 & \begin{array}{ccc} \displaystyle\left\{m_{b}\right\}\underset{1\leq b\leq N}{} &  \displaystyle\left\{f(\kappa_b)\right\}\underset{1\leq b\leq k}{} & 0 \end{array} \\ \hline \begin{array}{c} \displaystyle\left\{-m_{a}\right\}\underset{1\leq a\leq N}{} \\  \displaystyle\left\{-f(\kappa_a)\right\}\underset{1\leq a\leq k}{} \\ 0 \end{array} & M_{\rm even} \end{array}\right]\nonumber
 \end{eqnarray}
 for odd $N$ with the abbreviations
 \begin{eqnarray}\label{abbr}
  M_{ab}&=&\int_0^\infty dz_1dz_2\alpha(z_1)\alpha(z_2) z_1^{a-1}z_2^{b-1}\frac{z_1-z_2}{z_1+z_2},\\
  F_a(\kappa_b)&=&\int_0^\infty dz_1dz_2 \frac{\alpha(z_1)\alpha(z_2) z_1^{a-1}}{\kappa_b-z_2}\frac{z_1-z_2}{z_1+z_2},\nonumber\\
  G(\kappa_a,\kappa_b)&=&\int_0^\infty dz_1dz_2 \frac{\alpha(z_1)\alpha(z_2)}{(\kappa_a-z_1)(\kappa_b-z_2)}\frac{z_1-z_2}{z_1+z_2},\nonumber\\
  m_b&=&\int_0^\infty d z \alpha(z) z^{b-1},\nonumber\\
  f(\kappa_b)&=&\int_0^\infty \frac{d z \alpha(z)}{\kappa_b-z}.\nonumber
 \end{eqnarray}
 Note that the global sign is not important for the proof since we square the Pfaffian which yields determinants, i.e.
 \begin{eqnarray}\label{p2}
 \left(Z^{(N,{\rm B})}_{k|k}[\alpha](\kappa,\lambda)\right)^2=\left(\frac{1}{{\rm B}_{k|k}(\kappa;\lambda)}\right)^2\left\{\begin{array}{cl} \det M_{\rm even}, & N\ {\rm even}, \\ \det M_{\rm odd}, & N\ {\rm odd}.\end{array}\right. 
 \end{eqnarray}
  
In the next step we apply the simple relation
\begin{eqnarray}\label{simplerel}
 \frac{z_1-z_2}{z_1+z_2}=\frac{2z_1}{z_1+z_2}-1
\end{eqnarray}
and define the vectors
\begin{eqnarray}\label{vector-ev}
 \vec{x}_{\rm even}^T=\left[\begin{array}{ccc} \displaystyle\left\{m_{b}\right\}\underset{1\leq b\leq N}{} &  \displaystyle\left\{f(\kappa_b)\right\}\underset{1\leq b\leq k}{} & 0\ldots 0 \end{array}\right]
\end{eqnarray}
of dimension $N+2k$ and
\begin{eqnarray}\label{vector-odd}
 \vec{x}_{\rm odd}^T=\left[\begin{array}{c|ccc} -1 &\displaystyle\left\{m_{b}\right\}\underset{1\leq b\leq N}{} &  \displaystyle\left\{f(\kappa_b)\right\}\underset{1\leq b\leq k}{} & 0\ldots 0 \end{array}\right]
\end{eqnarray}
of dimension $N+2k+1$. Then we can use the following algebraic manipulation
\begin{eqnarray}\label{p3-ev}
 \det M_{\rm even}&=&\det M_{even}(1+\vec{x}_{\rm even}^TM_{\rm even}^{-1}\vec{x}_{\rm even})\\
 &=&\det M_{\rm even}\det(\mathbb{I}_{N+2k}+M_{\rm even}^{-1}\vec{x}_{\rm even}\vec{x}_{\rm even}^T)\nonumber\\
 &=&\det(M_{\rm even}+\vec{x}_{\rm even}\vec{x}_{\rm even}^T)\nonumber
\end{eqnarray}
for even $N$ and similarly for odd $N$,
\begin{eqnarray}\label{p3-odd}
 \det M_{\rm odd}&=&\det(M_{\rm odd}+\vec{x}_{\rm odd}\vec{x}_{\rm odd}^T).
\end{eqnarray}
Both relations are based on the fact that $M_{\rm even}$ as well as $M_{\rm odd}$ are antisymmetric, namely for any antisymmetric matrix $A$ and any vector $\vec{v}$ the expectation value $\vec{v}^TA\vec{v}$ vanishes.
 Defining the abbreviations
 \begin{eqnarray}\label{abbr2}
  \widehat{M}_{ab}&=&\int_0^\infty dz_1dz_2\alpha(z_1)\alpha(z_2) z_1^{a}z_2^{b-1}\frac{1}{z_1+z_2},\\
  \widehat{F}_a^{(1)}(\kappa_b)&=&\int_0^\infty dz_1dz_2 \frac{\alpha(z_1)\alpha(z_2) z_1^{a}}{\kappa_b-z_2}\frac{1}{z_1+z_2},\nonumber\\
  \widehat{F}_b^{(2)}(\kappa_a)&=&\int_0^\infty dz_1dz_2 \frac{\alpha(z_1)\alpha(z_2) z_1z_2^{b-1}}{\kappa_a-z_1}\frac{1}{z_1+z_2},\nonumber\\
  \widehat{G}(\kappa_a,\kappa_b)&=&\int_0^\infty dz_1dz_2 \frac{\alpha(z_1)\alpha(z_2)z_1}{(\kappa_a-z_1)(\kappa_b-z_2)}\frac{1}{z_1+z_2}.\nonumber
 \end{eqnarray}
 we use the identities~\eqref{p3-ev} and \eqref{p3-odd} such that
 \begin{eqnarray}\label{p4-ev}
 &&\left(Z^{(N,{\rm B})}_{k|k}[\alpha](\kappa,\lambda)\right)^2=2^N\left(\frac{1}{{\rm B}_{k|k}(\kappa;\lambda)}\right)^2\det\widehat{M}_{\rm even}\\
 &=&2^N\left(\frac{1}{{\rm B}_{k|k}(\kappa;\lambda)}\right)^2\nonumber\\
 &&\times\det\left[\begin{array}{ccc} \displaystyle\left\{\widehat{M}_{ab}\right\}\underset{1\leq a,b\leq N}{} & \displaystyle\left\{\widehat{F}_a^{(1)}(\kappa_b)\right\}\underset{1\leq b \leq k}{\underset{1\leq a\leq N}{}} & \displaystyle\left\{\lambda_b^{a-1}\right\}\underset{1\leq b \leq k}{\underset{1\leq a\leq N}{}} \\ \displaystyle\left\{\widehat{F}_b^{(2)}(\kappa_a)\right\}\underset{1\leq b \leq N}{\underset{1\leq a\leq k}{}} & \displaystyle\left\{\widehat{G}(\kappa_a,\kappa_b)\right\}\underset{1\leq a,b\leq k}{} & \displaystyle\left\{\frac{1}{\kappa_a-\lambda_b}\right\}\underset{1\leq a,b\leq k}{} \\ \displaystyle\left\{-\lambda_a^{b-1}\right\}\underset{1\leq b \leq N}{\underset{1\leq a\leq k}{}} & \displaystyle\left\{\frac{1}{\lambda_a-\kappa_b}\right\}\underset{1\leq a,b\leq k}{} & 0 \end{array}\right]\nonumber
 \end{eqnarray}
 for even $N$ and
 \begin{eqnarray}\label{p4-odd}
 \left(Z^{(N,{\rm B})}_{k|k}[\alpha](\kappa,\lambda)\right)^2&=&2^N\left(\frac{1}{{\rm B}_{k|k}(\kappa;\lambda)}\right)^2\det\widehat{M}_{\rm odd}\\
 &=&2^N\left(\frac{1}{{\rm B}_{k|k}(\kappa;\lambda)}\right)^2\det\left[\begin{array}{c|c} 1 & 0 \\ \hline \begin{array}{c} \displaystyle\left\{-2m_{a}\right\}\underset{1\leq a\leq N}{} \\  \displaystyle\left\{-2f(\kappa_a)\right\}\underset{1\leq a\leq k}{} \\ 0 \end{array} & \widehat{M}_{\rm even} \end{array}\right]\nonumber\\
 &=&2^N\left(\frac{1}{{\rm B}_{k|k}(\kappa;\lambda)}\right)^2\det\widehat{M}_{\rm even}\nonumber
 \end{eqnarray}
 for odd $N$.
 
 Now we can identify the integration variable $z_1$ in the kernels of the determinant~\eqref{p4-ev} with the integration variables $y_j$ in the proposition and the variable $z_2$ with the variables $x_j$. Then we can apply a generalized version of Andr\'eief's integration theorem~\cite{Andr,KG10} backwards and find
 \begin{eqnarray}
 \left(Z^{(N,{\rm B})}_{k|k}[\alpha](\kappa,\lambda)\right)^2&=&(-1)^{N(N-1)/2}2^N\int_{\mathbb R_+^{2N} }  \prod_{j=1}^{N}  \alpha(x_j) \alpha(y_j) y_j \,
  {\rm B}_{N|N}(x;-y)\nonumber\\
  &&\times\frac{{\rm B}_{k|N+k}(\kappa;x,\lambda)}{{\rm B}_{k|k}(\kappa;\lambda)}\frac{{\rm B}_{k|N+k}(\kappa;y,\lambda)}{{\rm B}_{k|k}(\kappa;\lambda)}(dx) (dy)\label{p4-identity.a}
 \end{eqnarray}
 independent of the fact whether $N$ is even or odd. In the last step we employ the definitions of the Cauchy-Vandermonde determinants~\eqref{Ber} and find the proposition for the case $l=k$. We can lift this restriction by extending the original integral for $k\neq l$ to the case $k=l$ and then taking the limit $\kappa_j\to\infty$ or $\lambda_j\to\infty$ for $j=\min\{k,l\}+1,\ldots,\max\{k,l\}$.

\section{Proof of proposition~\ref{Prop:det-structure-antisym}}\label{app:Prop2}

Choosing $L\in\mathbb{N}$ we define the matrix
  \begin{eqnarray}\label{mat-vec-def-a}
  \Delta M_L&=&\left[\Delta M_{ij}=\frac{1}{2}(\widehat{M}_{ij}-\widehat{M}_{ji})\right]_{1\leq i,j\leq L},
  \end{eqnarray}
  the vectors
  \begin{eqnarray}
  \vec{m}_L^T&=&(m_1,\ldots,m_L),\nonumber\\
  \Delta \vec{F}_L(\kappa) &=&\left(\frac{\widehat{F}_{1}^{(1)}(\kappa)-\widehat{F}_{1}^{(2)}(\kappa)}{2},\ldots,\frac{\widehat{F}_{L}^{(1)}(\kappa)-\widehat{F}_{L}^{(2)}(\kappa)}{2}\right),\nonumber\\
  \vec{\lambda}_{L}^T &=&\left(1,\lambda,\ldots,\lambda^{L-1}\right),\nonumber\\
  \vec{e}_{L+1}^T&=&(\overbrace{0,\ldots,0}^{L},1),\label{mat-vec-def-b}
  \end{eqnarray}
   and the scalar function
  \begin{eqnarray}\label{mat-vec-def-c}
  \Delta G(\kappa_1,\kappa_2)&=&\frac{1}{2}\left(\widehat{G}(\kappa_1,\kappa_2)-\widehat{G}(\kappa_2,\kappa_1)\right).
  \end{eqnarray}
  Then we can split the functions~\eqref{abbr2} in symmetric and anti-symmetric  parts,
  \begin{eqnarray}\label{redef}
   \widehat{M}_{ij}&=&\Delta M_{ij}+\frac{m_im_j}{2},\\
   \widehat{F}_j^{(1)}(\kappa)&=&\Delta F_j(\kappa)+\frac{ f(\kappa)m_j}{2},\nonumber\\
   \widehat{F}_j^{(2)}(\kappa)&=&-\Delta F_j(\kappa)+\frac{ f(\kappa)m_j}{2},\nonumber\\
   \widehat{G}(\kappa_1,\kappa_2)&=&\Delta \widehat{G}(\kappa_1,\kappa_2)+\frac{f(\kappa_1)f(\kappa_2)}{2}.\nonumber
  \end{eqnarray}
These relations follow from the particular form of the integrands, see Eqs.~\eqref{abbr} and \eqref{abbr2}.
We employ this splitting after we apply a generalized version of Andr\'eief's integration theorem~\cite{Andr,KG10}  to the four partition functions in the kernels of the determinant~\eqref{Cauchy-det-identity},
  \begin{eqnarray}
  Z^{(L+1,{\rm C})}_{1|0;1|0}[\alpha](\kappa_1;\kappa_2)&=&\det\left[\begin{array}{cc} \displaystyle\left\{\widehat{M}_{ij}\right\}\underset{1\leq i,j\leq L}{} & \displaystyle\left\{\widehat{F}_i^{(1)}(\kappa_1)\right\}\underset{1\leq i\leq L}{} \\ \displaystyle\left\{\widehat{F}_j^{(2)}(\kappa_2)\right\}\underset{1\leq j\leq L}{} & \widehat{G}(\kappa_1,\kappa_2) \end{array}\right]\nonumber\\
  &&\hspace*{-1cm}=\det\left[\begin{array}{c|c} \displaystyle\Delta M_L+\frac{\vec{m}_L\vec{m}_L^T}{2} & \displaystyle\Delta\vec{F}_L(\kappa_1)+\frac{f(\kappa_1)\vec{m}_L}{2} \\ \hline \displaystyle-\Delta\vec{F}_L^T(\kappa_2)+\frac{f(\kappa_2)\vec{m}_L^T}{2} & \displaystyle\Delta \widehat{G}(\kappa_1,\kappa_2)+\frac{f(\kappa_1)f(\kappa_2)}{2} \end{array}\right],\nonumber\\
  \frac{Z^{(L,{\rm C})}_{0|0;1|1}[\alpha](\kappa,\lambda)}{\kappa-\lambda}&=&\det\left[\begin{array}{ccc} \displaystyle\left\{\widehat{M}_{ij}\right\}\underset{1\leq i,j\leq L}{}  & \displaystyle\left\{\lambda^{i-1}\right\}\underset{1\leq i\leq L}{} \\ \displaystyle\left\{\widehat{F}_j^{(2)}(\kappa)\right\}\underset{1\leq j\leq L}{}  & \displaystyle\frac{1}{\kappa-\lambda} \end{array}\right]\nonumber\\
  &=&\det\left[\begin{array}{c|c} \displaystyle\Delta M_L+\frac{\vec{m}_L\vec{m}_L^T}{2} & \displaystyle\vec{\lambda}_L \\ \hline \displaystyle-\Delta\vec{F}_L^T(\kappa)+\frac{f(\kappa)\vec{m}_L^T}{2} & \displaystyle\frac{1}{\kappa-\lambda} \end{array}\right],\nonumber\\
  \frac{Z^{(L,{\rm C})}_{1|1;0|0}[\alpha](\kappa,\lambda)}{\kappa-\lambda}&=&\det\left[\begin{array}{ccc} \displaystyle\left\{\widehat{M}_{ij}\right\}\underset{1\leq i,j\leq L}{} & \displaystyle\left\{\widehat{F}_i^{(1)}(\kappa)\right\}\underset{1\leq i\leq L}{\ }  \\ \displaystyle\left\{\lambda^{j-1}\right\}\underset{1\leq j\leq L}{\ } & \displaystyle\frac{1}{\kappa-\lambda}  \end{array}\right]\nonumber\\
  &=&\det\left[\begin{array}{c|c} \displaystyle\Delta M_L+\frac{\vec{m}_L\vec{m}_L^T}{2} & \displaystyle\Delta\vec{F}_L(\kappa)+\frac{f(\kappa)\vec{m}_L}{2} \\ \hline \displaystyle\vec{\lambda}_L^T & \displaystyle\frac{1}{\kappa-\lambda} \end{array}\right],\nonumber\\
  Z^{(L-1,{\rm C})}_{0|1;0|1}[\alpha](\lambda_1;\lambda_2)&=&\det\left[\begin{array}{ccc} \displaystyle\left\{\widehat{M}_{ij}\right\}\underset{1\leq i,j\leq L}{} & \displaystyle\left\{\lambda_b^{i-1}\right\}\underset{1\leq i\leq L}{} \\ \displaystyle\left\{\lambda_a^{j-1}\right\}\underset{1\leq j\leq L}{} & 0 \end{array}\right]\nonumber\\
  &=&\det\left[\begin{array}{c|c} \displaystyle\Delta M_L+\frac{\vec{m}_L\vec{m}_L^T}{2} & \displaystyle\vec{\lambda}_{L,2} \\ \hline \displaystyle\vec{\lambda}_{L,1}^T & 0 \end{array}\right].\label{p5-1}
  \end{eqnarray}
The next steps only apply for the case $L\in2\mathbb{N}$ even since the antisymmetric matrix $\Delta M_L$ is only invertible in this case. For the case $L$ odd we have to modify this procedure. Exactly for this case we need the constant vector $\vec{e}_{L+1}$.

Let $L$ be even. Then the inverse of the matrix consisting of the moments of the Cauchy two-matrix ensemble reads
\begin{equation}\label{inversion}
 \left(\Delta M_L+\frac{\vec{m}_L\vec{m}_L^T}{2}\right)^{-1}=\Delta M_L^{-1}-\frac{1}{2}\Delta M_L^{-1}\vec{m}_L\vec{m}_L^T\Delta M_L^{-1}
\end{equation}
since we can perform a Taylor expansion in the dyadic matrix $\vec{m}_L\vec{m}_L^T$. This Taylor expansion is finite because $\vec{m}^T_L\Delta M_L^{-1}\vec{m}_L=0$ resulting from the asymmetry of $\Delta M_L$ which carries over to its inverse. Therefore we find
  \begin{eqnarray}
  \frac{Z^{(L+1,{\rm C})}_{1|0;1|0}[\alpha](\kappa_1;\kappa_2)}{Z^{(L,{\rm C})}_{0|0;0|0}[\alpha]}&=&\Delta \widehat{G}(\kappa_1,\kappa_2)+\frac{f(\kappa_1)f(\kappa_2)}{2}+\left(\Delta\vec{F}_L^T(\kappa_2)-\frac{f(\kappa_2)\vec{m}_L^T}{2}\right)\nonumber\\
  &&\hspace*{-1cm}\times\left(\Delta M_L^{-1}-\frac{1}{2}\Delta M_L^{-1}\vec{m}_L\vec{m}_L^T\Delta M_L^{-1}\right)\left(\Delta\vec{F}_L(\kappa_1)+\frac{f(\kappa_1)\vec{m}_L}{2}\right)\nonumber\\
  &=&-\frac{Z^{(L+1,{\rm C})}_{1|0;1|0}[\alpha](\kappa_2;\kappa_1)}{Z^{(L,{\rm C})}_{0|0;0|0}[\alpha]}\nonumber\\
  &&\hspace*{-1cm}+(f(\kappa_1)-\vec{m}_L^T\Delta M_L^{-1}\Delta \vec{F}_L(\kappa_1))(f(\kappa_2)-\vec{m}_L^T\Delta M_L^{-1}\Delta \vec{F}_L(\kappa_2)),\nonumber\\
  \frac{Z^{(L,{\rm C})}_{0|0;1|1}[\alpha](\kappa,\lambda)}{Z^{(L,{\rm C})}_{0|0;0|0}[\alpha](\kappa-\lambda)}&=&\frac{1}{\kappa-\lambda}+\left(\Delta\vec{F}_L^T(\kappa)-\frac{f(\kappa)\vec{m}_L^T}{2}\right)\nonumber\\
  &&\times\left(\Delta M_L^{-1}-\frac{1}{2}\Delta M_L^{-1}\vec{m}_L\vec{m}_L^T\Delta M_L^{-1}\right)\vec{\lambda}_L\nonumber\\
  &=&\frac{Z^{(L,{\rm C})}_{1|1;0|0}[\alpha](\kappa,\lambda)}{Z^{(L,{\rm C})}_{0|0;0|0}[\alpha](\kappa-\lambda)}\nonumber\\
  &&+(f(\kappa)-\vec{m}_L^T\Delta M_L^{-1}\Delta \vec{F}_L(\kappa))\vec{\lambda}_L^T\Delta M_L^{-1}\vec{m},\nonumber\\
  \frac{Z^{(L-1,{\rm C})}_{0|1;0|1}[\alpha](\lambda_1;\lambda_2)}{Z^{(L,{\rm C})}_{0|0;0|0}[\alpha]}&=&-\vec{\lambda}_{L,1}^T\Delta M^{-1}\vec{\lambda}_{L,2}+\frac{1}{2}\vec{m}_L^T\Delta M_L^{-1}\vec{\lambda}_{L,1}\vec{m}_L^T\Delta M_L^{-1}\vec{\lambda}_{L,2}\nonumber\\
  &=&-\frac{Z^{(L-1,{\rm C})}_{0|1;0|1}[\alpha](\lambda_2;\lambda_1)}{Z^{(L,{\rm C})}_{0|0;0|0}[\alpha]}+\vec{m}_L^T\Delta M_L^{-1}\vec{\lambda}_{L,1}\vec{m}_L^T\Delta M_L^{-1}\vec{\lambda}_{L,2}.\nonumber\\ \label{p5-2}
  \end{eqnarray}
 These relations prove the proposition for the case $L$ even and $k+l$ even. The reason for this is that we can define the vector
 \begin{equation}
 \vec{v}=\left[\begin{array}{c} \displaystyle\left\{f(\kappa_a)-\vec{m}_L^T\Delta M_L^{-1}\Delta \vec{F}_L(\kappa_j)\right\}_{1\leq j\leq k} \\ \displaystyle\left\{\vec{m}_L^T\Delta M_L^{-1}\vec{\lambda}_{L,j}\right\}_{1\leq j\leq l} \end{array}\right]
 \end{equation}
such that the partition function is
 \begin{equation}
 Z^{(L-l+k,{\rm C})}_{k|l;k|l}[\alpha](\kappa,\lambda;\kappa,\lambda)=\frac{Z^{(L,{\rm C})}_{0|0;0|0}[\alpha]}{{\rm B}_{k|l}^2(\kappa;\lambda)}\det\left(K+\frac{1}{2}\vec{v}\vec{v}^T\right)=\frac{Z^{(L,{\rm C})}_{0|0;0|0}[\alpha]}{{\rm B}_{k|l}^2(\kappa;\lambda)}\det K,\label{p5-3}
 \end{equation}
 where $K$ is the remaining antisymmetric matrix in the determinant~\eqref{det-antisym-even}. We extent this result to $L$ even and $k+l$ odd by introducing an auxiliary variable $\lambda_{l+1}$ in the partition function 
 \begin{equation}
 Z^{(L-l+k,{\rm C})}_{k|l;k|l}[\alpha](\kappa,\lambda;\kappa,\lambda)=\lim\limits_{\lambda_{l+1}\to\infty}\frac{Z^{(L-l+k,{\rm C})}_{k|l+1;k|l+1}[\alpha](\kappa,\lambda;\kappa,\lambda)}{\lambda_{l+1}^{2(L-l+k)}}.
 \end{equation}
 
 Let us come to the other case where $L\in2\mathbb{N}_0+1$ odd. Then the antisymmetric matrix $\Delta M_L$ is not invertible anymore. Therefore we extend the determinants~\eqref{p5-1} by the vector $\vec{e}_{L+1}$,
  \begin{eqnarray}
  &&Z^{(L+1,{\rm C})}_{1|0;1|0}[\alpha](\kappa_1;\kappa_2)\nonumber\\
  &=&-\det\left[\begin{array}{c|c|c} \displaystyle\Delta M_{L+1}+\frac{\vec{m}_{L+1}\vec{m}_{L+1}^T}{2} & \displaystyle\Delta\vec{F}_{L+1}(\kappa_1)+\frac{f(\kappa_1)\vec{m}_{L+1}}{2} & \vec{e}_{L+1} \\ \hline \displaystyle-\Delta\vec{F}_{L+1}^T(\kappa_2)+\frac{f(\kappa_2)\vec{m}_{L+1}^T}{2} & \displaystyle\Delta \widehat{G}(\kappa_1,\kappa_2)+\frac{f(\kappa_1)f(\kappa_2)}{2} & 0 \\ \hline \vec{e}_{L+1}^T & 0 & 0 \end{array}\right],\nonumber\\
  &&\frac{Z^{(L,{\rm C})}_{0|0;1|1}[\alpha](\kappa,\lambda)}{\kappa-\lambda}\nonumber\\
  &=&-\det\left[\begin{array}{c|c|c} \displaystyle\Delta M_{L+1}+\frac{\vec{m}_{L+1}\vec{m}_{L+1}^T}{2} & \displaystyle\vec{\lambda}_{L +1} & \vec{e}_{L+1} \\ \hline \displaystyle-\Delta\vec{F}_{L+1}^T(\kappa)+\frac{f(\kappa)\vec{m}_{L+1}^T}{2} & \displaystyle\frac{1}{\kappa-\lambda} & 0 \\ \hline \vec{e}_{L+1}^T & 0 & 0 \end{array}\right],\nonumber\\
  &&\frac{Z^{(L,{\rm C})}_{1|1;0|0}[\alpha](\kappa,\lambda)}{\kappa-\lambda}\nonumber\\
  &=&-\det\left[\begin{array}{c|c|c} \displaystyle\Delta M_{L+1}+\frac{\vec{m}_{L+1}\vec{m}_{L+1}^T}{2} & \displaystyle\Delta\vec{F}_{L+1}(\kappa)+\frac{f(\kappa)\vec{m}_{L+1}}{2} & \vec{e}_{L+1} \\ \hline \displaystyle\vec{\lambda}_{L+1}^T & \displaystyle\frac{1}{\kappa-\lambda} & 0\\ \hline \vec{e}_{L+1}^T & 0 & 0 \end{array}\right],\nonumber\\
  &&Z^{(L-1,{\rm C})}_{0|1;0|1}[\alpha](\lambda_1;\lambda_2)\nonumber\\
  &=&-\det\left[\begin{array}{c|c|c} \displaystyle\Delta M_{L+1}+\frac{\vec{m}_{L+1}\vec{m}_L^T}{2} & \displaystyle\vec{\lambda}_{L+1,2} & \vec{e}_{L+1} \\ \hline \displaystyle\vec{\lambda}_{L+1,1}^T & 0 & 0\\ \hline \vec{e}_{L+1}^T & 0 & 0 \end{array}\right].\label{p5-4}
  \end{eqnarray}
  Now the antisymmetric matrix $\Delta M_{L+1}$ is invertible and we can derive relations analogous to those in Eq.~\eqref{p5-2},
  \begin{eqnarray}
  \frac{Z^{(L+1,{\rm C})}_{1|0;1|0}[\alpha](\kappa_1;\kappa_2)}{Z^{(L,{\rm C})}_{0|0;0|0}[\alpha]}&=&-\frac{Z^{(L+1,{\rm C})}_{1|0;1|0}[\alpha](\kappa_2;\kappa_1)}{Z^{(L,{\rm C})}_{0|0;0|0}[\alpha]}\nonumber\\
  &&+2\vec{e}_{L+1}^T\Delta M_{L+1}^{-1}\Delta \vec{F}_{L+1}(\kappa_1)\vec{e}_{L+1}^T\Delta M_{L+1}^{-1}\Delta \vec{F}_{L+1}(\kappa_2)\nonumber\\
  \frac{Z^{(L,{\rm C})}_{0|0;1|1}[\alpha](\kappa,\lambda)}{Z^{(L,{\rm C})}_{0|0;0|0}[\alpha](\kappa-\lambda)}&=&\frac{Z^{(L,{\rm C})}_{1|1;0|0}[\alpha](\kappa,\lambda)}{Z^{(L,{\rm C})}_{0|0;0|0}[\alpha](\kappa-\lambda)}\nonumber\\
  &&+2\vec{e}_{L+1}^T\Delta M_{L+1}^{-1}\Delta \vec{F}_{L+1}(\kappa)\vec{e}_{L+1}\Delta M_{L+1}^{-1}\vec{\lambda}_{L+1},\nonumber\\
  \frac{Z^{(L-1,{\rm C})}_{0|1;0|1}[\alpha](\lambda_1;\lambda_2)}{Z^{(L,{\rm C})}_{0|0;0|0}[\alpha]}&=&-\frac{Z^{(L-1,{\rm C})}_{0|1;0|1}[\alpha](\lambda_2;\lambda_1)}{Z^{(L,{\rm C})}_{0|0;0|0}[\alpha]}\nonumber\\
  &&+2\vec{e}_{L+1}\Delta M_{L+1}^{-1}\vec{\lambda}_{L+1,1}\vec{e}_{L+1}\Delta M_{L+1}^{-1}\vec{\lambda}_{L+1,2}.\label{p5-5}
  \end{eqnarray}
  Hence we have to define a new vector
 \begin{equation}
 \vec{v}=\left[\begin{array}{c} \displaystyle\left\{\vec{e}_{L+1}^T\Delta M_{L+1}^{-1}\Delta \vec{F}_{L+1}(\kappa_j)\right\}_{1\leq j\leq k} \\ \displaystyle\left\{\vec{e}_{L+1}\Delta M_{L+1}^{-1}\vec{\lambda}_{L+1,j}\right\}_{1\leq j\leq l} \end{array}\right]
 \end{equation} 
 with which one can do a calculation similar to Eq.~\eqref{p5-3} and so proving the proposition for the case $L$ odd and $k+l$ even. The case for $k+l$ odd can be performed exactly in the same way as for $L$ even which finishes the proof.

\providecommand{\bysame}{\leavevmode\hbox to3em{\hrulefill}\thinspace}
\providecommand{\MR}{\relax\ifhmode\unskip\space\fi MR }
\providecommand{\MRhref}[2]{%
  \href{http://www.ams.org/mathscinet-getitem?mr=#1}{#2}
}
\providecommand{\href}[2]{#2}

\end{document}